\documentclass[12pt]{article}
\usepackage[centertags]{amsmath}
\usepackage{amsfonts}
\usepackage{amssymb}
\usepackage{amsthm}
\usepackage{newlfont}
\usepackage{epsfig}
\usepackage{amscd}

\newcommand{\beq}{\begin{equation}}
\newcommand{\eeq}{\end{equation}}
\newcommand{\ba}{\begin{array}}
\newcommand{\ea}{\end{array}}
\newcommand{\bea}{\begin{eqnarray}}
\newcommand{\eea}{\end{eqnarray}}

\begin{document}

\begin{center}
{\large \sc \bf Differential reductions  of the Kadomtsev-Petviashvili
equation and associated higher dimensional nonlinear PDEs. }

\vskip 15pt

{\large 
%P. M. Santini$^{1,\S}$ and 
A. I. Zenchuk
%$^{2,\S}$ 
}

\vskip 8pt

%{\it
%$^1$ Dipartimento di Fisica, Universit\`a di Roma "La Sapienza" %and\\
%Istituto Nazionale di Fisica Nucleare, Sezione di Roma 1\\
%Piazz.le Aldo Moro 2, I-00185 Roma, Italy}

%\smallskip

{\it %$^2$ 
Institute of Chemical Physics RAS,
Acad. Semenov av., 1
Chernogolovka,
Moscow region
142432,
Russia}

\smallskip

\vskip 5pt

%$^{\S}$
e-mail:  {
%\tt paolo.santini@roma1.infn.it , 
zenchuk@itp.ac.ru }

\vskip 5pt

{\today}

\end{center}

\begin{abstract}
We represent an algorithm allowing one to construct  new classes of
partially integrable  multidimensional nonlinear partial differential equations (PDEs) starting with the special type of solutions to the (1+1)-dimensional hierarchy of nonlinear PDEs linearizable by the matrix Hopf-Cole substitution (the B\"urgers hierarchy).
 We derive  examples of four-dimensional nonlinear matrix  PDEs together with they scalar and three-dimensional reductions. Variants of the  Kadomtsev-Petviashvili type and Korteweg-de Vries  type equations are represented among them. Our algorithm is based on the combination of two Frobenius type reductions and  special differential reduction imposed on the matrix fields of  integrable PDEs. It is shown that the derived four-dimensional nonlinear PDEs admit arbitrary functions of two variables in their solution spaces which clarifies the integrability degree of these PDEs.
\end{abstract}

%%%%%%%%%%%%%%%%
\section{Introduction}

The problem of construction  of such   multidimensional nonlinear partial differential equations (PDEs) which are either completely integrable or, at least, possess a big manifold of particular solutions is very attractive problem of integrability theory. First of all, one should emphasize several  remarkable classical works  regarding the completely integrable models such as \cite{GGKM} (where the Korteweg-de Vries equation (KdV) has been first time studied), \cite{ZSh1,ZSh2} (where the dressing method for a big class of soliton and instanton equations has been formulated),  \cite{OSTT} (where the so-called Sato approach to the integrability is applied to Kadomtsev-Petviashvili equation (KP)). A big class of the first order systems of  quasilinear PDEs is integrated in \cite{ts1,dn} using the generalized hodograph method.  However, most of the above integrable nonlinear PDEs are (2+1)- and/or (1+1)-dimensional, excepting the Selfdual-type PDEs \cite{Ward,BZ} and the equations assotiated with commuting vector fields  \cite{Krichever,TT,DMT,KAR,MS1,MS2}. 

In this paper we use modification of the algorithm  represented in \cite{Z5,ZS_2} allowing one to construct  new multidimensional partially integrable nonlinear PDEs. It is shown in \cite{ZS_2}  that Frobenious reduction of the matrix fields  of the nonlinear PDEs integrable  either by the Hoph-Cole substitution \cite{Hopf} ($C$-integrable PDE \cite{Calogero,Calogero2,Calogero3,Calogero4,Calogero5,Calogero6}) or by the  method of characteristics \cite{Whitham,SZ} ($Ch$-integrable PDE) leads to one  of two big classes of the  nonlinear PDEs integrable by the inverse spectral transform method  (ISTM)  \cite{ZMNP, CD, AC, Konop} ($S$-integrable PDEs \cite{Calogero}). These classes are ($a$) soliton equations, such as KdV \cite{GGKM,KdV}, the Nonlinear Schr\"odinger equation (NLS) \cite{ZS_NLS}, the Kadomtsev-Petviashvili equation (KP) \cite{KP}, the Deavi-Stewartson equation (DS) \cite{DS}, and ($b$) instanton equations, such as the Self-dual  Yang-Mills equation (SDYM) \cite{Ward,BZ}. 

The natural question is whether the Frobenius reduction (or its modification) can be used for construction  new types of integrable (or at least partially integrable)  systems starting with any known integrable system or this method works only for derivation of soliton and instanton  PDEs from $C$- and $Ch$-integrable ones?  

At first glance the answer is negative. In fact,  one can  verify  that  Frobenius reduction applied to such   matrix $S$-integrable PDE as the GL($N$) SDYM, the $N$-wave equation and the KP  produces the same $S$-integrable PDE. 
However, there is a method to generate new higher dimensional partially integrable systems of nonlinear PDEs  using 
Frobenius type reduction after the appropriate differential reduction imposed on the matrix fields of the above $S$-integrable  nonlinear PDEs.

Such combination of reductions has been already used in \cite{Z3}. It is shown there  that GL($N$) SDYM supplemented by the pair of reductions, namely, differential reduction relating certain blocks  of the matrix field and Frobenius reduction of these blocks, produces a new five-dimensional system of matrix nonlinear PDEs (with three-dimensional solution space), whose scalar reduction results in the  nonlinear PDE assotiated with commuting vector fields
\cite{MS1,MS2}.

Following the strategy of ref.\cite{Z3},
 we consider an algorithm  for  construction new partially  integrable PDEs starting with the matrix KP (although  this algorithm may be applied to any $S$-integrable model).  We will derive two representative of matrix systems whose scalar reductions yield the  following four-dimensional equations:
\begin{eqnarray}\label{nl_ex1}\label{nl_ex1_sc}
&&
\left(v_{x\tau}-\frac{1}{2}v_{yt_2}-\frac{1}{2}\nu_0 v_{xx}\right)_y - \frac{1}{2}v_{xxx} +v_{yyy}v_x -v_y v_{xyy}=0,\\\label{nl_ex2}\label{KP_type}
&&
v_{xt_3} -\frac{3}{4} v_{t_2t_2}-\frac{1}{4} v_{xxxx} +\frac{3}{2} \Big(v_{yt_2} v_{x} -v_{t_2} v_{xy} -(v_{x}\partial_y^{-1}v_{xx} )_x\Big)=0.
 \end{eqnarray}
Another reduction, 
\begin{eqnarray}\label{red}
v_{t_2}=0,
\end{eqnarray}
reduces these PDEs into the following three-dimensional  ones
\begin{eqnarray}
\label{ex_3dim1}
&&
\left(v_{x\tau}-\frac{1}{2}\nu_0 v_{xx}\right)_y - \frac{1}{2}v_{xxx} +  v_{yyy}v_x -v_y v_{xyy}=0,\\\label{ex_3dim2}
&&
v_{t_3}-\frac{1}{4} v_{xxx} -\frac{3}{2} v_x \partial_y^{-1} v_{xx} =0. 
\end{eqnarray}
Here we take $\tau$ and $t_3$ as evolutionary parameters (times) while $x$, $y$ and $t_2$ are taken as space parameters. Since the linear parts of eqs.(\ref{nl_ex2}) and (\ref{ex_3dim2}) coincide with the linear parts of the KP and the KdV respectively, eqs.(\ref{nl_ex2}) and (\ref{ex_3dim2})
may be treated as new variants of KP- and KdV-type equations respectively. The feature of these equations is that the derivative with respect to  $y$ appears only in the nonlinear parts. For this reason, these equations may be referred to as dispersionless ones.

Remember, that the KP originates from  the (1+1)-dimensional $C$-integrable B\"urgers hierarchy due to the Frobenius reduction \cite{ZS_2}. Thus the complete set of transformations leading to eqs.(\ref{nl_ex1}) and (\ref{nl_ex2}) is following (see also Fig.1). 
We start with the $C$-integrable B\"urgers hierarchy of nonlinear PDEs with independent variables $x$ and $t_n$, $n=2,3,\dots$. Frobenius reduction of this hierarchy \cite{ZS_2} yields the  proper hierarchy of discrete chains of nonlinear PDEs, which is equivalent to the chains obtained in the Sato approach to the integrability of (2+1)-dimensional  KP \cite{OSTT}.
%(2+1)-dimensional  $S$-integrable PDE \cite{OSTT} 
%(which is KP in our case). 
The later may be derived as an intermediate result of our algorithm  after  eliminating all extra fields using combination of the first and the second representatives of the constructed discrete hierarchy. Next, apply the differential reduction 
introducing one more independent variable $y$. This step yields a three-dimensional  system of  nonlinear PDEs, i.e. dimensionality coincides with that of KP. Finally, the Frobenius type reduction applied to the matrix fields of the latter nonlinear  PDEs results in a four-dimensional system of matrix PDEs whose scalar versions yield eqs.(\ref{nl_ex1},\ref{nl_ex2}). It will be shown that the derived four-dimensional nonlinear PDEs may  not be completely integrated by our method  because the  available  solution spaces to them are restricted.

%%%%%%%%%%%%%%%%%%%%%%%%%%%%%%%%%%%%%%%%%%%%%%%%%%%%%%%%%%%%%%%%%%%%%
\vskip 20pt
\begin{center}
\mbox{ \epsfxsize=15cm \epsffile{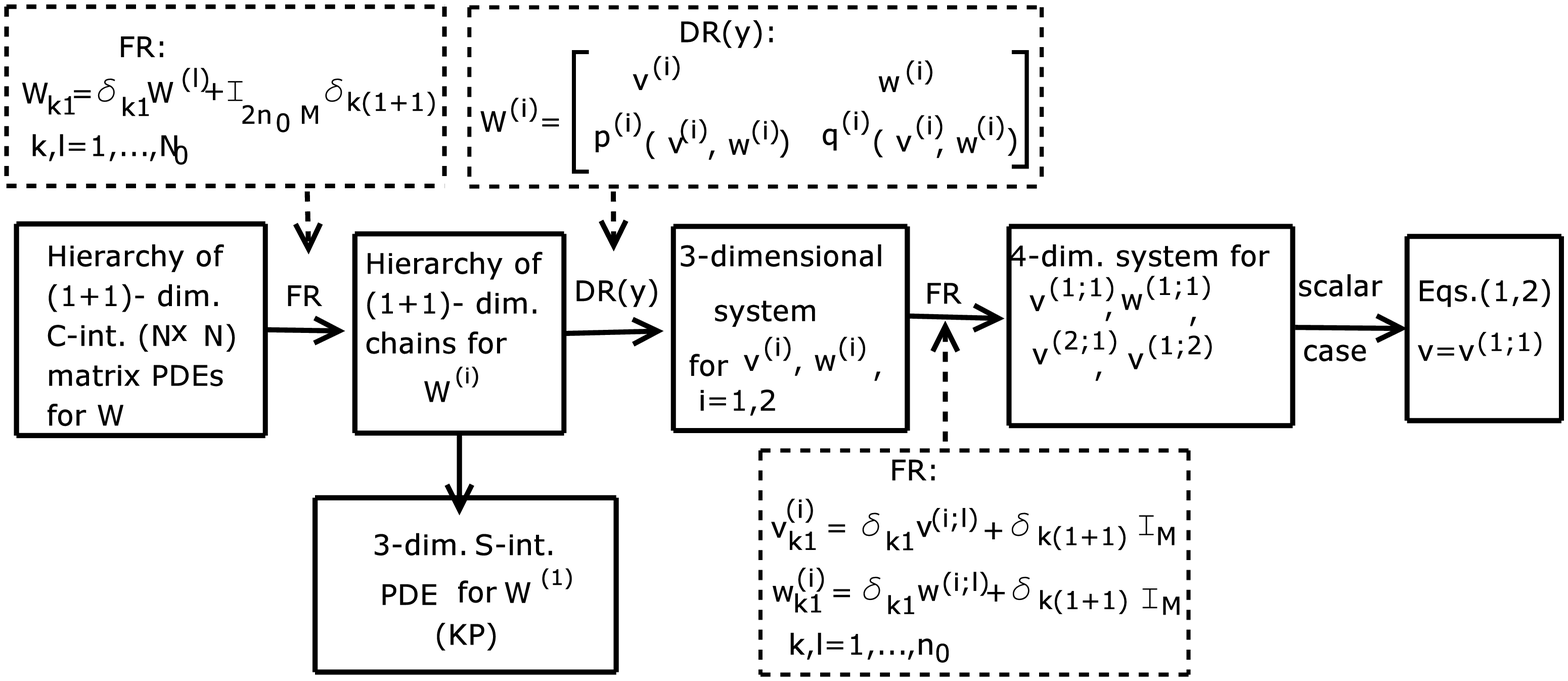}}
\end{center}
Fig.1 {\small The chain of transformations from the 
(1+1)-dimensional $C$-integrable B\"urgers hierarchy to the four-dimensional  PDEs (\ref{nl_ex1}) and (\ref{nl_ex2}). Here $N=2 N_0 n_0 M$,  $W^{(i)}$ are $2 n_0 M\times 2 n_0 M$ matrix fields, $v^{(i)}$, $w^{(i)}$, $p^{(i)}$ and $q^{(i)}$ are $n_0 M\times n_0M$ blocks,  $I_{2 n_0 M}$ and $I_M$ are $2 n_0 M$- and $M$-dimensional identity matrices respectively. Blocks  $p^{(i)}$ and $q^{(i)}$ are defined by eqs.(\ref{qp_i}). Here $N_0$, $n_0$ and $M$ are arbitrary positive integer parameters.}
\vskip 20pt

%%%%%%%%%%%%%%%%%%%%%%%%%%%%%%%%%%%%%%%%%%%%%%%%%%%%%%%%%%%%%%%%%%%%%%

%We may guaranty the presence of two arbitrary functions of two 
%variables therein. 

This paper is organized as follows. In Sec.\ref{Section:KP} we 
briefly recall some results of \cite{ZS_2} and derive the discrete chains of nonlinear PDEs produced by the (1+1)-dimensional B\"urgers hierarchy (with independent variables $x$ and $t_n$, $n=2,3$) supplemented by the Frobenius reduction. Using a few equations of theses chains we  eliminate all extra fields and derive the matrix KP.  In Sec.\ref{Section:DR} we suggest the special type differential reduction imposed on the blocks of the $2n_0M\times 2n_0M$ matrix fields of the above chains. This reduction introduces a new independent variables $y$ and $\tau$ and allows one to generate $n_0M\times n_0M$ three-dimensional  matrix PDEs. Frobenius type reduction of the above PDEs results in the four-dimensional systems of $M\times M$ matrix PDEs, Sec.\ref{Section:FTR}. Scalar versions of these PDEs result in eqs.(\ref{nl_ex1}) and (\ref{nl_ex2}) which, in particular, may be reduced to eqs.(\ref{ex_3dim1}) and (\ref{ex_3dim2}) respectively.  Solution spaces to the nonlinear PDEs derived in Secs.\ref{Section:DR} and \ref{Section:FTR} will be studied in Sec.\ref{Section:solutions}. The obstacles to the complete integrability of eqs.(\ref{nl_ex1}), (\ref{nl_ex2}), (\ref{ex_3dim1}) and  (\ref{ex_3dim2}) as  well as the obstacles to their integrability by the ISTM are briefly discussed in Sec.\ref{Section:obstacle}. Conclusions are given in Sec.\ref{Section:conclusions}.

%%%%%%%%%
\section{Relation between the (1+1)-dimensional $C$-integrable B\"urgers hierarchy and   the matrix KP}
\label{Section:KP}
\subsection{The (1+1)-dimensional $C$-integrable B\"urgers hierarchy} 
Here we use the B\"urgers hierarchy as a simplest example of $C$-integrable hierarchies  linearizable by the matrix Hopf-Cole substitution. Namely, let $W$ be solution of the following linear algebraic matrix equation:
\begin{eqnarray}\label{lin_chi}\label{chiW}
&&
\hat\chi_x=W\hat\chi,\\\label{lin_t}
&&
D_{t_n}\hat \chi=\partial_x^n\hat\chi,\;\;
D_{t_n}\hat \chi\equiv \hat \chi_{t_n} +\hat \chi \Lambda^{(n)},\;\;n=2,3,\dots.
\end{eqnarray}
Here $\hat \chi$ and $W$ are $2N_0n_0 M\times 2N_0n_0 M$ matrix functions, $\Lambda^{(n)}$ are  $2N_0n_0 M\times 2N_0n_0 M$ commuting constant matrices.  To anticipate, the matrices $\Lambda^{(n)}$ are introduced in order to establish the reductions eliminating derivatives with respect to  $t_n$ from the nonlinear PDEs, see, for example,  reduction (\ref{red}). The compatibility conditions of eqs.
(\ref{lin_chi}) and (\ref{lin_t}) yield  the linearizable B\"urgers hierarchy. We will need only the first and the second representatives of this hierarchy below, i.e. $n=2,3$ in eqs.(\ref{lin_t}):
\begin{eqnarray}\label{lin_t2}
&&
W_{t_2}-W_{xx}-2 W_x W=0,\\\label{lin_t3}
&&
W_{t_3}-W_{xxx}-3 W_{xx} W -3W_x(W_x+W^2)=0.
\end{eqnarray}

%%%%%%%%%%%%%
\subsection{The Frobenius reduction and assotiated chains of nonlinear PDEs}
Introduce the Frobenius  reduction \cite{ZS_2}: 
\begin{eqnarray}\label{ww}
W&=&\left(
\begin{array}{ccccc}
W^{(1)} &W^{(2)} &\cdots& W^{(N_0-1)} & W^{(N_0)}\cr
I_{2Mn_0} &0_{2Mn_0}&\cdots & 0_{2Mn_0}& 0_{2Mn_0}\cr
0_{2Mn_0} &I_{2Mn_0}&\cdots & 0_{2Mn_0}& 0_{2Mn_0}\cr
\cdots&\cdots&\cdots&\cdots&\cdots\cr
0_{2Mn_0} &0_{2Mn_0}&\cdots & I_{2Mn_0}&0_{2Mn_0}
\end{array}
\right)
,
\end{eqnarray}
where $I_{J}$ and $0_{J}$ are $J\times J$ identity and zero matrices respectively, $W^{(i)}$ are $2 n_0M\times 2n_0M$ matrix functions.
Substituting eq.(\ref{ww}) into eq.(\ref{chiW}) 
we obtain the  following block structure of $\hat\chi$:
\begin{eqnarray}\label{chi}
\hat\chi&=&\left(
\begin{array}{ccc}
\chi^{(1)} &\cdots & \chi^{(N_0)}\cr
\partial_x^{-1}\chi^{(1)} &\cdots & \partial_x^{-1}\chi^{(N_0)}\cr
\cdots&\cdots&\cdots\cr
\partial_x^{-N_0+1}\chi^{(1)} &\cdots & \partial_x^{-N_0+1}\chi^{(N_0)}
\end{array}
\right),
\end{eqnarray}
where  $\chi^{(i)}$ are $2 n_0M\times 2n_0M$ matrix functions.
Thus, eqs.(\ref{lin_chi}) and (\ref{lin_t}) may be written as  follows:
\begin{eqnarray}\label{lin_KP}
&&
\chi^{(m)}_x(\vec x)=\sum_{i= 1}^{N_0} W^{(i)}(\vec x) \partial_x^{-i+1}\chi^{(m)}(\vec x),\\\label{t_KP}
&&
\tilde D_{t_n}\chi^{(m)}= \partial_x^n\chi^{(m)},\;\;\chi=(\chi^{(1)} \;\; \dots \;\; \chi^{(N_0)}),
\\\label{D_x}
&&
\tilde D_{t_n}\chi^{(m)} = \chi^{(m)}_{t_n} + \chi^{(m)} 
\Lambda^{(n;m)},\;\;\Lambda^{(n)}={\mbox{diag}}(
\Lambda^{(n;1)},\dots,\Lambda^{(n;N_0)}),\\\nonumber
&&
m=1,\dots,N_0,
\;\;n=2,3,\dots,
\end{eqnarray}
where   $\Lambda^{(n;m)}$ are some $2 n_0 M\times 2 n_0 M$ commuting constant matrices. The meaning of the integer parameters $n_0$ and $M$ is clarified in Fig.1. 
The  compatibility conditions of eqs.(\ref{lin_KP}) and (\ref{t_KP})
(with $n=2,3$) yield the following discrete chains:
\begin{eqnarray}\label{Burgers_ch}
&&
W^{(n)}_{t_{2}} -  W^{(n)}_{xx}- 2 W^{(n+1)}_x
  -2  W^{(1)}_x W^{(n)} =0 ,
\\\label{Burgers-3_ch}
&&
W^{(n)}_{t_{3}} -W^{(n)}_{xxx} -3\Big( 
 W^{(1)}_{xx} W^{(n)}+ W^{(n+1)}_{xx} +  W^{(1)}_x (W^{(1)} W^{(n)} + W^{(n+1)} + W^{(n)}_x)+\\\nonumber
&&
W^{(2)}_xW^{(n)}+W^{(n+2)}_{x}\Big) =0 ,  \;\;\; n=1,\dots, N_0,\\\nonumber
&&
W^{(l)}=0,\;\;l>N_0.
\end{eqnarray} 
Alternatively, these chains may be derived substituting  eq.(\ref{ww}) into eqs.(\ref{lin_t2}) and (\ref{lin_t3}).

%%%%%%%%%%%%%%
\subsection{Matrix KP and its scalar reduction}
The matrix KP is represented by the system of three equations involving fields $W^{(i)}$, $i=1,2,3$:  eq.(\ref{Burgers_ch}) with $n=1,2$ and  eq.(\ref{Burgers-3_ch}) with $n=1$.  Eliminating $W^{(2)}$ and $W^{(3)}$ from this system one gets the following nonlinear PDE for the field $ W^{(1)}$:
\begin{eqnarray}\label{matrixKP}
\Big(
 W^{(1)}_{t_3} -\frac{1}{4} W^{(1)}_{xxx} -\frac{3}{2}(W^{(1)}_x)^2
 \Big)_x+\frac{3}{2} [W^{(1)}_x,W^{(1)}_{t_2}]-\frac{3}{4} W^{(1)}_{t_2t_2}=0 ,
\end{eqnarray}
where square parenthesis mean matrix commutator.
 In the scalar case this equation  reduces to the following one, $u\equiv W^{(1)}$:
 \begin{eqnarray}\label{KP}
 \Big(
 u_{t_3} -\frac{1}{4} u_{xxx} -\frac{3}{2}u_x^2
 \Big)_x-\frac{3}{4} u_{t_2t_2}=0 
 \end{eqnarray}
which is the scalar potential KP.

%%%%%%%%%%%%%
\section{Differential reduction of the matrix KP}
\label{Section:diff_reduction}
\label{Section:DR}
Let
matrices $\chi^{(j)}$ have the following block structure:
\begin{eqnarray}\label{dif_red}
&&
\chi^{(j)}=\left(\begin{array}{cc}
\Psi^{(2j-1)}&\Psi^{(2j)}\cr
\Psi_y^{(2j-1)}&\Psi_y^{(2j)}\cr
\end{array}\right),\;\;\;
j=1,\dots,N_0,
\end{eqnarray}
where  $\Psi^{(m)}$ are $n_0M\times n_0M$ matrix functions.
We introduce  $y$-dependence of $\Psi^{(m)}$ by the following second order PDE:
\begin{eqnarray}
\label{Psi_xx}\label{g_r1_3_2}\label{g_r2}\label{x}
{\cal{E}}^{(0)}&:=&\Psi^{(m)}_{yy} = a \Psi^{(m)}_x+\nu \Psi_y^{(m)}  + \mu\Psi^{(m)},\;\;m=1,\dots,2N_0,\;\;
\end{eqnarray}
Here  $a$, $\nu$ and $\mu$ are  $n_0M\times n_0M$ diagonal constant matrices.
The block structure of $\chi^{(j)}$  (\ref{dif_red}) suggests us the  relevant  block structure of $W^{(j)}$:
\begin{eqnarray}\label{ww_i}
&&
W^{(j)}=\left(\begin{array}{cc}
w^{(j)} & v^{(j)}\cr
p^{(j)} & q^{(j)}
\end{array}\right),\;\;\;j=1,\dots,N_0,\\\nonumber
&&
v^{(l)}=w^{(l)}=p^{(l)}=q^{(l)}=0_{Mn_0},\;\;l>N_0,
\end{eqnarray}
where $w^{(j)}$, $v^{(j)}$, $q^{(j)}$ and $p^{(j)}$ are $ n_0M\times n_0 M$ matrix functions.
Now matrix  $2 n_0  M\times 2 n_0 M$   equations (\ref{lin_KP}) may be written as two $ n_0M \times 2N_0 n_0 M $ equations:
\begin{eqnarray}\label{g_r1}\label{b1}
&&
{\cal{E}}^{(1)}:=\Psi_x =\sum_{i=1}^{N_0}\Big(
v^{(i)} \partial_x^{-i+1}\Psi_y +w^{(i)} \partial_x^{-i+1}\Psi\Big), \\\label{b2}
&&
{\cal{E}}^{(2)}:=\Psi_{xy}  =\sum_{i=1}^{N_0}\Big(
q^{(i)} \partial_x^{-i+1}\Psi_y +p^{(i)} \partial_x^{-i+1}\Psi\Big),
\end{eqnarray}
where
\begin{eqnarray}\label{Psi^}
&&
\Psi=(\Psi^{(1)},\dots,\Psi^{(2 N_0)}).
\end{eqnarray}
The compatibility condition of  eqs.(\ref{b1}) and (\ref{b2}),
\begin{eqnarray}
{{\cal{E}}^{(1)}_y=\cal{E}}^{(2)},
\end{eqnarray}
yields the expressions for $p^{(j)}$ and $q^{(j)}$ in terms of $v^{(j)}$ and $w^{(j)}$:
\begin{eqnarray}\label{qp_i}
&&
p^{(j)}=
w^{(j)}_y+v^{(j)}\mu + v^{(j+1)}  a+v^{(1)} a w^{(j)} ,\;\;\;q^{(j)}= v^{(j)}_y +v^{(j)} \nu + w^{(j)} + v^{(1)}  a v^{(j)}.
\end{eqnarray}
 Thus only two blocks of $W^{(j)}$ are independent, i.e. $w^{(j)}$ and $v^{(j)}$.
Matrix equation (\ref{g_r1}) may be considered  as the uniquely solvable  system of $2N_0$  linear $n_0 M \times n_0 M$  matrix algebraic equations 
for the matrix functions $v^{(i)}$ and $w^{(i)}$, $i=1,\dots,N_0$, while eq.(\ref{b2}) is the consequence of  eq.(\ref{g_r1}).

Eqs.(\ref{t_KP})  yield:
\begin{eqnarray}\label{t_r1}\label{t}
&&
\hat D_{t_n}\Psi^{(m)} = \partial_x^n\Psi^{(m)},\;\;\hat D_{t_n}\Psi^{(m)}\equiv \Psi^{(m)}_{t_n} + \Psi^{(m)} \tilde \Lambda^{(n;m)},\;\;m=1,\dots,2N_0,\\\nonumber
&&
\Lambda^{(n;j)}={\mbox{diag}}(\tilde \Lambda^{(n;2j-1)},\tilde  \Lambda^{(n;2j)}),\;\;j=1,\dots,N_0,\;\;
n=2,3,\dots,
\end{eqnarray}
where $\tilde\Lambda^{(n,m)}$ are $n_0M\times n_0M$ commuting constant matrices.
Eq.(\ref{x}) allows us to introduce one more set of parameters $\tau_n$ as follows:
\begin{eqnarray}\label{tau}
&&
 \Psi^{(m)}_{\tau_{n}} =\partial_x^n \Psi^{(m)}_y  ,\;\;m=1,\dots,2N_0,\;\;n=1,2,\dots.
\end{eqnarray}

Before proceeding further we  select the following three equations out of  system (\ref{t},\ref{tau}):
\begin{eqnarray}\label{t2}
&&
\hat D_{t_2}\Psi^{(m)}= \Psi^{(m)}_{xx},\\\label{tau1}
&&
\Psi^{(m)}_\tau\equiv \Psi^{(m)}_{\tau_1}=  \Psi^{(m)}_{xy},\\\label{t3}
&&
\hat D_{t_3}\Psi^{(m)} =  \Psi^{(m)}_{xxx}.
\end{eqnarray}
We also assume that $a$, $\nu$ and $\mu$ are scalars, i.e.
\begin{eqnarray}\label{a_nu_mu}
a= I_{n_0M},\;\;\nu=\nu_0 I_{n_0M},\;\;\mu=\mu_0 I_{n_0M}.
\end{eqnarray}
In order to derive the nonlinear PDEs for $v^{(i)}$ and $w^{(i)}$, we must consider  the 
 conditions providing the compatibility of eq.(\ref{b1}) and eqs.(\ref{t2}-\ref{t3}):
\begin{eqnarray}\label{comp1}
&&
\hat D_{t_2}{\cal{E}}^{(1)}-{\cal{E}}^{(1)}_{xx}=0\;\;
\Rightarrow\\\nonumber
&&
\sum_{i=1}^{N_0}E^{(1i)}_1 \partial_x^{-i+1}D_y\Psi
+\sum_{i=1}^{N_0}E^{(1i)}_0 \partial_x^{-i+1}\Psi
  =0,\;\;\;n=1,2,\dots,\\\label{comp2}
&&
{\cal{E}}^{(1)}_\tau-{\cal{E}}^{(1)}_{xy}=0\;\;\Rightarrow\\\nonumber
&&
\sum_{i=1}^{N_0}E^{(2i)}_1 \partial_x^{-i+1}\Psi_y
+\sum_{i=1}^{N_0}E^{(2i)}_0 \partial_x^{-i+1}\Psi
  =0,\;\;\;n=1,2,\dots,\\\label{comp3}
&&
\hat D_{t_3}{\cal{E}}^{(1)}-{\cal{E}}^{(1)}_{xxx}=0\;\;
\Rightarrow\\\nonumber
&&
\sum_{i=1}^{N_0}E^{(3i)}_1 \partial_x^{-i+1}\Psi_y
+\sum_{i=1}^{N_0}E^{(3i)}_0 \partial_x^{-i+1}\Psi
  =0,\;\;\;n=1,2,\dots.
\end{eqnarray}
These equations  generate  the following chains of  nonlinear PDEs  for $v^{(i)}$ and $w^{(i)}$, $i=1,\dots,N_0$:
%\begin{eqnarray}\label{ch_1}
%E^{(1i)}_1=0,\;\;\;E^{(1i)}_0=0,\\\label{ch_2}
%E^{(2i)}_1=0,\;\;\;E^{(2i)}_0=0,\\\label{ch_3}
%E^{(3i)}_1=0,\;\;\;E^{(3i)}_0=0.
%\end{eqnarray}
%Explicite forms of these equations read:
\begin{eqnarray}\label{nl_ch_1}
E^{(1i)}_1&\equiv& v^{(i)}_{t_2} - v^{(i)}_{xx} - 2  v^{(i+1)}_{x} -2 (\nu_0 v^{(1)}_x +w^{(1)}_x+ v^{(1)}_x v^{(1)}) v^{(i)} -2  v^{(1)}_x (w^{(i)} + v^{(i)}_y)=0,\\\label{nl_ch_11}
E^{(1i)}_0&\equiv&w^{(i)}_{t_2}-w^{(i)}_{xx} -2 w^{(i+1)}_x -2 (w^{(1)}_x + v^{(1)}_x v^{(1)}) w^{(i)} -2 v^{(1)}_x (w^{(i)}_y +v^{(i+1)} + \mu_0 v^{(i)}) =0,
\end{eqnarray}
\begin{eqnarray}
\label{nl_ch_2}
E^{(2i)}_1&\equiv& v^{(i)}_{\tau} - v^{(i)}_{xy} -   v^{(i+1)}_{y}-\nu_0 v^{(i)}_x - w^{(i)}_x-(\nu_0 v^{(1)}_y + w^{(1)}_y+ v^{(1)}_x + v^{(1)}_y v^{(1)}) v^{(i)}- \\\nonumber
&&
v^{(1)}_y (w^{(i)} + v^{(i)}_y)=0,\\\label{nl_ch_21}
E^{(2i)}_0&\equiv&
w^{(i)}_{\tau} - w^{(i)}_{xy} -   w^{(i+1)}_{y}-v^{(i+1)}_{x}-\mu_0 v^{(i)}_x -(w^{(1)}_y +v^{(1)}_x + v^{(1)}_y v^{(1)} )w^{(i)}-
\\\nonumber
&&
 v^{(1)}_y (v^{(i+1)} +w^{(i)}_y+\mu_0 v^{(i)})=0 
,
\end{eqnarray}
\begin{eqnarray}
\label{nl_ch_3}
E^{(3i)}_1&\equiv& v^{(i)}_{t_3} - v^{(i)}_{xxx} -   3v^{(i+1)}_{xx}-3v^{(i+2)}_{x}-3 \Big(  w^{(2)}_x
+ 
w^{(1)}_{xx}+v^{(1)}_x v^{(2)}+
 v^{(1)}_x w^{(1)}_y+(v^{(1)}_x)^2+\\\nonumber
&&
v^{(2)}_x v^{(1)}+
w^{(1)}_x w^{(1)}+
v^{(1)}_{xx} v^{(1)}+v^{(1)}_x  v^{(1)} w^{(1)}+v^{(1)}_x  w^{(1)} v^{(1)}+v^{(1)}_x  v^{(1)}_y v^{(1)}+
w^{(1)}_x  (v^{(1)})^2+\\\nonumber
&&
v^{(1)}_x  (v^{(1)})^3
+
\nu_0( v^{(1)}_{xx}+v^{(1)}_x w^{(1)}+ v^{(1)}_x v^{(1)}_y+ v^{(2)}_x
+ w^{(1)}_x v^{(1)}+\nu_0 v^{(1)}_x  v^{(1)}+2 v^{(1)}_{x} (v^{(1)})^2)+\\\nonumber
&&
\mu_0 v^{(1)}_x v^{(1)}
\Big) v^{(i)} 
-3 v^{(1)}_x\Big(
w^{(i+1)}  +v^{(i+1)}_y +w^{(i)}_x+
v^{(i)}_{xy}+v^{(1)}v^{(i+1)}+v^{(1)}v^{(i)}_x+\\\nonumber
&&
w^{(1)}v^{(i)}_y+
v^{(1)}_y w^{(i)}+v^{(1)}_y v^{(i)}_y+
(v^{(1)})^2 w^{(i)} +(v^{(1)})^2 v^{(i)}_y +w^{(1)} w^{(i)}
+\nu_0( v^{(i+1)} +v^{(i)}_x+\\\nonumber
&&
v^{(1)}w^{(i)}+v^{(1)}v^{(i)}_y)
\Big)-3 \Big(v^{(2)}_x (w^{(i)}+v^{(i)}_y)+w^{(1)}_x v^{(i)}_x+w^{(1)}_x v^{(i+1)}+v^{(1)}_{xx} v^{(i)}_y 
+v^{(1)}_{xx}w^{(i)}+\\\nonumber
&&
w^{(1)}_x(v^{(1)} w^{(i)} +v^{(1)} v^{(i)}_y) 
\Big)=0,\\\label{nl_ch_31}
E^{(3i)}_0&\equiv& w^{(i)}_{t_3} - w^{(i)}_{xxx} -   3w^{(i+1)}_{xx}-3w^{(i+2)}_{x} -3 \Big(w^{(1)}_{xx} +w^{(2)}_x +
 v^{(2)}_x v^{(1)}+w^{(1)}_x w^{(1)}+v^{(1)}_{xx} v^{(1)}+
 \\\nonumber
 &&
v^{(1)}_{x} (v^{(1)}w^{(1)}+w^{(1)}v^{(1)}+v^{(1)}_yv^{(1)}+(v^{(1)})^3+ v^{(2)}+ w^{(1)}_y+v^{(1)}_x)
+
w^{(1)}_x (v^{(1)})^2
+\\\nonumber
 &&
 \nu_0  v^{(1)}_x (v^{(1)})^2 
+\mu_0 v^{(1)}_x v^{(1)} 
\Big) w^{(i)}-3 v^{(1)}_x\Big(
v^{(i+2)}+w^{(i+1)}_y+v^{(i+1)}_x+w^{(i)}_{xy}
+ v^{(1)}w^{(i+1)}+
\\\nonumber
 &&
v^{(1)}w^{(i)}_x+w^{(1)}v^{(i+1)}+
w^{(1)}w^{(i)}_y+v^{(1)}_yv^{(i+1)}+v^{(1)}_yw^{(i)}_y
+(v^{(1)})^2(v^{(i+1)}+w^{(i)}_y)+\\\nonumber
 &&
\nu_0 (v^{(1)}v^{(i+1)}+ v^{(1)}w^{(i)}_y)+
\mu_0 (v^{(1)}_yv^{(i)}
+
 w^{(1)}v^{(i)}
+v^{(i)}_x
+(v^{(1)})^2v^{(i)}+ v^{(i+1)}+
\\\nonumber
 &&
\nu_0 v^{(1)}v^{(i)})
\Big)-3 v^{(2)}_x (v^{(i+1)} + w^{(i)}_y +
\mu_0 v^{(i)})
-3 w^{(1)}_x \Big(
w^{(i+1)} +w^{(i)}_x +v^{(1)}(v^{(i+1)}+\\\nonumber
 &&
 w^{(i)}_y+\mu_0 v^{(i)})
\Big) -3 v^{(1)}_{xx} (v^{(i+1)} + w^{(i)}_y+\mu_0v^{(i)})
\end{eqnarray}

In addition, we must to take into account the compatibility condition of eqs.(\ref{g_r2}) and (\ref{g_r1}),
\begin{eqnarray}\label{constr1}
{\cal{E}}^{(1)}_{yy}&=&{\cal{E}}^{(1)}_x+\nu_0{\cal{E}}^{(1)}_y+\mu_0{\cal{E}}^{(1)} \;\;\Rightarrow\;\;
\sum_{i=0}^{N_0}(\tilde E^{(i)}_1\partial_x^{-i+1}\Psi_y+\tilde E^{(i)}_0\partial_x^{-i+1}\Psi) =0
,
\end{eqnarray}
which gives us the following system of non-evolutionary  nonlinear chains:
\begin{eqnarray}\label{nl4}
\tilde E^{(i)}_1&:=&v^{(i)}_{yy} + 2 w^{(i)}_y -  v^{(i)}_x +\nu_0 v^{(i)}_y  +
2   v^{(1)}_y  v^{(i)} =0,
\\\label{nl4_1}
\tilde E^{(i)}_0&:=& w^{(i)}_{yy} - \nu_0 w^{(i)}_y -  w^{(i)}_x +2  v^{(i+1)}_y +2 \mu_0 v^{(i)}_y +
2   v^{(1)}_y  w^{(i)} =0,\;\;\;i=1,\dots,N_0.
\end{eqnarray}
%Of course, the system (\ref{nl_ch_1}-\ref{nl_ch_31},\ref{nl4}) %may be derived directly from the system %(\ref{Burgers_ch},\ref{Burgers-3_ch}) using reduction %(\ref{ww_i}). However, the derivation of this system as the %compatibility condition of the linear system %(\ref{t},\ref{g_r2},\ref{g_r1}) is more illustrative.

One can show that  
the derived chains (\ref{nl_ch_1}-\ref{nl_ch_3},\ref{nl4},\ref{nl4_1}) generate three  compatible three-dimensional systems of nonlinear PDEs as follows:
\begin{eqnarray}\label{dif_1}
E^{(11)}_i=0,\;\;\;\tilde E^{(1)}_i=0,&&i=0,1,\;\; {\mbox{fields $v^{(j)}$, $w^{(j)}$, $j=1,2$,}}\\\nonumber
&& {\mbox{ independent variables $t_2,x,y$}}, \\\label{dif_2}
E^{(21)}_i=0,\;\;\;\tilde E^{(1)}_i=0,&&i=0,1,\;\;{\mbox{fields $v^{(j)}$, $w^{(j)}$, $j=1,2$,}} \\\nonumber
&& {\mbox{independent variables $\tau,x,y$}},\\\label{dif_3}
E^{(31)}_i=0,\;\;\;\tilde E^{(k)}_i=0,&&i=0,1,\;\;k=1,2,\;\;{\mbox{fields $v^{(j)}$, $w^{(j)}$, $j=1,2,3$,}}\\\nonumber
&&{\mbox{ independent variables $t_3,x,y$}}.
\end{eqnarray}
We do not represent these equations explicitely since they are intermediate equations in our algorithm.

%%%%%%%%%%%%%%
\section{The Frobenius type reduction, new four-dimensional systems of matrix nonlinear PDEs and their scalar reductions}
\label{Section:FTR}
Three systems (\ref{dif_1}-\ref{dif_3}) represent three commuting flows with times $t_2$, $\tau$ and $t_3$ respectively. 
In this section we follow the strategy of ref.\cite{ZS_2} and show that  hierarchy of  nonlinear chains (\ref{nl_ch_1}-\ref{nl_ch_31},\ref{nl4},\ref{nl4_1})
%any two  three-dimensional systems out of %(\ref{dif_1}-\ref{dif_3}) 
supplemented by the Frobenius type reduction
% generate single  four-dimensional system of nonlinear  matrix %PDEs. More precisely, it will be demonstrated that any two %equations out of three equations $E^{(11)}_1=0$, $E^{(21)}_1=0$, %$E^{(31)}_1=0$ supplemented by the eq.
%$\tilde E^{(1)}_1$ and by the Frobenius type reductions
 of the matrix fields $v^{(i)}$ and $w^{(i)}$ generates the hierarchy of  four-dimensional systems of nonlinear PDEs, see eqs.(\ref{nl_ch_1_fr_1_newvar}-\ref{nl4_fr_1_newvar}) and text thereafter.

This is possible due to the 
remarkable property of   chains of nonlinear PDEs (\ref{nl_ch_1}-\ref{nl_ch_31},\ref{nl4},\ref{nl4_1}). Namely, these chains  admit  the following Frobenious type reduction:
\begin{eqnarray}\label{G_F}
 &&
 v^{(i)}=\{v^{(i;kl)},\;\;k,l=1,\dots,n_0\},\;\;
  w^{(i)}=\{w^{(i;kl)},\;\;k,l=1,\dots,n_0\},
  \\\nonumber
  &&
  v^{(i;kl)}=\delta_{k1} v^{(i;l)} +
  \delta_{k (l+n_1(i))} I_M ,\\\nonumber
  &&
  w^{(i;kl)}=\delta_{k 1} w^{(i;l)} +
  \delta_{k(l+n_2(i))} I_M  ,\;\;i=1,\dots,N_0,\\\nonumber
  &&
  v^{(i;l)}=w^{(i;l)}=0_M,\;\;l>n_0,\;\;\forall\;i,
  \end{eqnarray}
where  $n_0$ is an arbitrary positive integer parameter, $n_1(i)$ and $n_2(i)$ are arbitrary positive integer functions of positive integer argument,
$v^{(i;l)}$ and $w^{(i;l)}$ are $M\times M$ matrix fields.
In particular, if $n_1(i)=n_2(i)=1$, then this reduction becomes Frobenius one \cite{ZS_2}, which is shown in Fig.1.

Eqs.(\ref{G_F}) require the following 
block-structures of the functions $\Psi^{(l)}$ and of the constant matrices $\tilde\Lambda^{(i;l)}$:
\begin{eqnarray} &&
\Psi^{(l)}=\{\Psi^{(l;nm)},\;n,m=1,\dots,n_0\},\\\nonumber
&&
\tilde\Lambda^{(i;l)}=\{\Lambda^{(i;l;m)},\;m=1,\dots,n_0\},\;\;l=1,\dots,2N_0,\;\;i=2,3.
\end{eqnarray}
Here $\Psi^{(l;nm)}$ are $M\times M$ matrix functions and $\Lambda^{(i;l;m)}$ are  $M\times M$ commuting constant matrices.
In turn, eq.(\ref{b1}) reduces to the following one:
\begin{eqnarray}\label{Psi_ex}
\Psi^{(l;nm)}_x&=&
\sum_{i=1}^{N_0}\sum_{j=1}^{n_0}\left[\Big(\delta_{n1}v^{(i;j)} +
\delta_{n(j+n_1(i))} \Big)
\partial_x^{-i+1}  \Psi^{(l;jm)}_y+
\Big(\delta_{n1} w^{(i;j)} + \right.\\\nonumber
&&\left.
\delta_{n(j+n_2(i))}  \Big)
\partial_x^{-i+1}\Psi^{(l;jm)}\right],\;\;l=   1,\dots,2N_0,\;\;n,m=1,\dots,n_0,
\end{eqnarray}
while eqs.(\ref{t2}-\ref{t3},\ref{x})  yield 
\begin{eqnarray}\label{t2_ex}
&&
{\cal{D}}_{t_2}\Psi^{(l;nm)} =\Psi^{(l;nm)}_{xx} ,\\\label{tau_ex}
&&
\Psi^{(l;nm)}_\tau =\Psi^{(l;nm)}_{xy},\\ \label{t3_ex}
&&
{\cal{D}}_{t_3}\Psi^{(l;nm)} =\Psi^{(l;nm)}_{xxx}
\\\label{x_ex}
&&
 \Psi^{(l;nm)}_{yy}=  \Psi^{(l;nm)}_x +\nu_0  \Psi^{(l;nm)}_y + \mu_0\Psi^{(l;nm)},
\\\nonumber
&&
{\cal{D}}_{t_i}\Psi^{(l;nm)}=\Psi^{(l;nm)}_{t_i} +  \Psi^{(l;nm)}\Lambda^{(i;l;m)},\;\;i=2,3,\\\nonumber
&&
l=1,\dots,2N_0,\;\;n,m=1,\dots,n_0.
\end{eqnarray}
Then the chains of  nonlinear PDEs (\ref{nl_ch_1}-\ref{nl_ch_31},\ref{nl4},\ref{nl4_1}) get the following block structures:
\begin{eqnarray}
&&
E^{(ni)}_m=\{E^{(ni;l)}_m\delta_{k1},\;\;k,l,=1,\dots,n_0\}=0,\;\;n=1,2,3,
\\\nonumber
&&
\tilde E^{(i)}_m=\{\tilde E^{(i;l)}_m\delta_{k1},\;\;k,l,=1,\dots,n_0\}=0,\\\nonumber
&&m=0,1,\;\;\;i=1,\dots,N_0.
\end{eqnarray} 
In particular, nonlinear chains (\ref{nl_ch_1},\ref{nl_ch_11}) and (\ref{nl4},\ref{nl4_1}) with
Frobenious type reduction (\ref{G_F})  generate the following chains with two discrete indexes:
\begin{eqnarray}\label{nl_ch_1_fr}
 E^{(1i;l)}_1&\equiv& v^{(i;l)}_{t_2} - v^{(i;l)}_{xx} - 2  v^{(i+1;l)}_{x} -2 (\nu_0 v^{(1;1)}_x +w^{(1;1)}_x+ v^{(1;1)}_x v^{(1;1)}) v^{(i;l)} -
 2  v^{(1;1)}_x (w^{(i;l)} +\\\nonumber
 &&
  v^{(i;l)}_y)-
 2 \Big(\nu_0 v^{(1;l+n_1(i))}_x +w^{(1;l+n_1(i))}_x+ 
 v^{(1;1+n_1(1))}_x v^{(i;l)}+ 
v^{(1;1)}_x v^{(1;l+n_1(i))}+
\\\nonumber
 &&
v^{(1;l+n_1(1)+n_1(i))}_x
 +  v^{(1;l+n_2(i))}_x\Big)=0,
 \\\label{nl_ch_11_fr}
E^{(1i;l)}_0&\equiv&w^{(i;l)}_{t_2}-w^{(i;l)}_{xx} -2 w^{(i+1;l)}_x -2 (w^{(1;1)}_x + v^{(1;1)}_x v^{(1;1)}) w^{(i;l)} -2 v^{(1;1)}_x (w^{(i;l)}_y +v^{(i+1;l)} +\\\nonumber
&& \mu_0 v^{(i;l)}) -2 \Big(w^{(1;l+n_1(i))}_x +
v^{(1;1+n_1(1))}_x w^{(i;l)}+ v^{(1;1)}_x v^{(1;l+n_2(i))}+v^{(1;l+n_1(1)+n_2(i))}_x+
\\\nonumber
&&
v^{(1;l+n_1(i+1))}_x +\mu_0 v^{(1;l+n_1(i))}_x \Big) =0,
\end{eqnarray}
\begin{eqnarray}\label{nl4_fr}
\tilde E^{(i;l)}_1&:=&v^{(i;l)}_{yy} + 2  w^{(i;l)}_y -  v^{(i;l)}_x +\nu_0 v^{(i;l)}_y  +
2   v^{(1;1)}_y  v^{(i;l)} +2  v^{(i;l+n_1(i))}_y=0,\\
\label{nl4_1_fr}
\tilde E^{(i;l)}_0&:=& w^{(i;l)}_{yy} - \nu_0 w^{(i;l)}_y -  w^{(i;l)}_x +2  v^{(i+1;l)}_y +2 \mu_0 v^{(i;l)}_y +
2   v^{(1;1)}_y  w^{(i;l)} +
2   v^{(i;l+n_2(i))}_y =0,
\end{eqnarray}
where $i=1,\dots, N_0$, $l=1,\dots,n_0$. 
Now our goal is to write the complete system of PDEs for some fields $v^{(i;j)}$ and/or $w^{(n;m)}$ which would be independent on both parameters $N_0$ and $n_0$ (in the spirit of the Sato theory \cite{OSTT}). This may be done if, along with system (\ref{nl_ch_1_fr}-\ref{nl4_1_fr}), one  involves
the discrete chains of PDEs generated by either eqs.(\ref{nl_ch_2},\ref{nl_ch_21}) or eqs.(\ref{nl_ch_3},\ref{nl_ch_31}) together with the Frobenious type reduction (\ref{G_F}).
For instance,  system (\ref{nl_ch_2},\ref{nl_ch_21}) yields
\begin{eqnarray}
 \label{nl_ch_2_fr}
E^{(2i;l)}_1&\equiv& v^{(i;l)}_{\tau} - v^{(i;l)}_{xy} -   v^{(i+1;l)}_{y}-\nu_0 v^{(i;l)}_x - w^{(i;l)}_x-(\nu_0 v^{(1;1)}_y + w^{(1;1)}_y+ v^{(1;1)}_x + \\\nonumber
&&
v^{(1;1)}_y v^{(1;1)}) v^{(i;l)}- 
v^{(1;1)}_y (w^{(i;l)} + v^{(i;l)}_y)-(\nu_0 v^{(i;l+n_1(i))}_y + w^{(i;l+n_1(i))}_y+ v^{(i;l+n_1(i))}_x + \\\nonumber
&&
v^{(1;1+n_1(1))}_y v^{(i;l)}+ 
v^{(1;1)}_y v^{(1;l+n_1(i))}+v^{(1;l+ n_1(1)+n_1(i))}_y)-v^{(1;l+n_2(i))}_y=0,\\\label{nl_ch_22_fr}
 E^{(2i;l)}_0&\equiv&
 w^{(i;l)}_{\tau} - w^{(i;l)}_{xy} -   w^{(i+1;l)}_{y}-v^{(i+1;l)}_{x}-\mu_0 v^{(i;l)}_x -(w^{(1;1)}_y +v^{(1;1)}_x + v^{(1;1)}_y v^{(1;1)} )w^{(i;l)}-
\\\nonumber
&&
 v^{(1;1)}_y (v^{(i+1;l)} +w^{(i;l)}_y+\mu_0 v^{(i;l)})-w^{(1;l+n_2(i))}_y-v^{(1;l+n_2(i))}_x-
 (v^{(1;1+n_1(1))}_y w^{(i;l)}+
 \\\nonumber
 &&
  v^{(1;1)}_y v^{(1;l+n_2(i))}+v^{(1;l+n_1(1)+n_2(i))}_y)-v^{(1;l+n_1(i+1) )}_y-v^{(1;l+n_1(i) )}_y\mu_0=0.
\end{eqnarray}
The discrete chains generated by eqs. (\ref{nl_ch_3}) and (\ref{nl_ch_31}) are very cumbersome. However, it will be shown that  we  need only equation (\ref{nl_ch_3}) with $i=1$ for derivation of the complete system of nonlinear PDEs. The later equation  may be simplified after  elimination $v^{(2)}_x$, $v^{(3)}_x$ and $w^{(2)}_x$ using eq.(\ref{nl_ch_1}) with $i=1,2$ and eq.(\ref{nl_ch_11}) with $i=1$. This  simplification is  reasonable because, in any case, we are going to use chain (\ref{nl_ch_1_fr}) (produced  by eqs.(\ref{nl_ch_1})) in order to derive the complete system of nonlinear PDEs.
%, as has been indicated above. 
%(\ref{KP_type}) in the Example 2 of this section. 
One gets in result: 
\begin{eqnarray}\label{nl_ch_3_mod}
&&
4 v^{(1)}_{t_3} -v^{(1)}_{xxx} - 3 v^{(1)}_{xt_2} -
6 v^{(2)}_{t_2} - 6 \Big(w^{(1)}_{t_2}+ v^{(1)}_{t_2} v^{(1)}  + (v^{(1)}_x)^2+
 \nu_0 v^{(1)}_{t_2}  \Big)v^{(1)} -\\\nonumber
&&
6 v^{(1)}_x\Big(w^{(1)}_x +v^{(1)}_{xy}+v^{(1)} v^{(1)}_x +\nu_0 v^{(1)}_x\Big)-
6 v^{(1)}_{t_2} (w^{(1)}+ v^{(1)}_y) -6 w^{(1)}_x v^{(1)}_x=0.
\end{eqnarray}
After the  Frobenius reduction, this equation generates the following discrete chain:
\begin{eqnarray}\label{nl_ch_3_mod_fr}
&&
4 v^{(1;l)}_{t_3} -v^{(1;l)}_{xxx} - 3 v^{(1;l)}_{xt_2} -
6 v^{(2;l)}_{t_2} - 6 \Big(w^{(1;1)}_{t_2}+ v^{(1;1)}_{t_2} v^{(1;1)}  + (v^{(1;1)}_x)^2+
 \nu_0 v^{(1;1)}_{t_2}  \Big)v^{(1;l)} -\\\nonumber
&&
6 v^{(1;1)}_x\Big(w^{(1;l)}_x +v^{(1;l)}_{xy}+v^{(1;1)} v^{(1;l)}_x +\nu_0 v^{(1;l)}_x\Big)-
6 v^{(1;1)}_{t_2} (w^{(1;l)}+ v^{(1;l)}_y) -6 w^{(1;1)}_x v^{(1;l)}_x
-\\\nonumber
&& 6 \Big(w^{(1;l+n_1(1))}_{t_2}+ v^{(1;l+2 n_1(1))}_{t_2} +v^{(1;1+n_1(1))}_{t_2}v^{(1;l)}+ v^{(1;1)}_{t_2}v^{(1;l+n_1(1))}  + v^{(1;1)}_xv^{(1;l+n_1(1))}_x+\\\nonumber
&&
 \nu_0 v^{(1;l+n_1(1))}_{t_2} +
 v^{(1;1+n_1(1))}_x  v^{(1;l)}_x+
 v^{(1;l+n_2(1))}_{t_2} \Big)=0.
\end{eqnarray} 
%Finally, eq.(\ref{nl4}) after Frobenius reduction  yields:
%\begin{eqnarray}\label{nl4_fr}
%\tilde E^{(i;l)}_1&:=&v^{(i;l)}_{yy} + 2  w^{(i;l)}_y -  %v^{(i;l)}_x +\nu_0 v^{(i;l)}_y  +
%2   v^{(1;1)}_y  v^{(i;l)} +2  v^{(i;l+n_1(i))}_y=0.
%\end{eqnarray}

%The explicite forms of the equations $E^{(ni;l)}_m=0$, $n,m=1,2$ %and 
%$\tilde E^{(i;l)}_m=0$, $m=1,2$ as well as $E^{(31;l)}_m=0$ %after some simplifications (equations $\tilde E^{(31;l)}_m=0$)  %are given in Appendix, see eqs.(\ref{dif_1}-\ref{dif_3}). 

\iffalse
As a consequence, systems of nonlinear PDEs (\ref{dif_1}-\ref{dif_3}) generate the following compatible chains of nonlinear PDEs, $l=1,\dots,n_0$:
\begin{eqnarray}\label{dif_1_F}
E^{(11;l)}_i=0,\;\;\;\tilde E^{(1;l)}_i=0,&&i=0,1,\;\; {\mbox{fields $v^{(i;l)}$, $w^{(i;l)}$, $i=1,2$,}}\\\nonumber
&& {\mbox{ independent variables $t_2,x,y$}}, \\\label{dif_2_F}
E^{(21;l)}_i=0,\;\;\;\tilde E^{(1;l)}_i=0,&&i=0,1,\;\;{\mbox{fields $v^{(i;l)}$, $w^{(i;l)}$, $i=1,2$,}} \\\nonumber
&& {\mbox{independent variables $\tau,x,y$}},\\\label{dif_3_F}
E^{(31;l)}_i=0,\;\;\;\tilde E^{(j;l)}_i=0,&&i=0,1,\;\;j=1,2,\;\;{\mbox{fields $v^{(i;l)}$, $w^{(i;l)}$, $i=1,2,3$,}}\\\nonumber
&&{\mbox{ independent variables $t_3,x,y$}}.
\end{eqnarray}
\fi

Having chains of equations  
(\ref{nl_ch_1_fr}-\ref{nl_ch_22_fr},\ref{nl_ch_3_mod_fr}) we would like to exhibit the complete system of nonlinear PDEs independent on both $N_0$ and $n_0$ taking a few equations out of these chains.
After some examination of these chains 
 we conclude that
it is enough to take the equations $E^{(k1;1)}_1$, $k=1,2$, $\tilde E^{(31;1)}_1$ and $\tilde E^{(1;1)}_1$.
Introducing the new fields
\begin{eqnarray}\label{newvar}
&&
v=v^{(1;1)},\;\;q= v^{(1;1+n_1(1))} +w^{(1;1)},\\
\nonumber
&&
p=v^{(2;1)} + w^{(1;1+n_1(1))} +v^{(1;1+n_2(1))}+v^{(1;1+2n_1(1))}+\nu_0 v^{(1;1+n_1(1))},
\end{eqnarray}
we write these equations as follows:
\begin{eqnarray}\label{nl_ch_1_fr_1_newvar}
 E^{(11;1)}_1&\equiv& v_{t_2} - v_{xx} - 2  p_{x} -2 (\nu_0 v_x +q_x+ v_x v) v -
 2  v_x (q +  v_y)=0,\\\label{nl_ch_2_fr_1_newvar}
E^{(21;1)}_1&\equiv& v_{\tau} - v_{xy} -   p_{y}-\nu_0 v_x - q_x-(\nu_0 v_y + q_y+ v_x + 
v_y v) v- 
v_y (q + v_y)=0
,
\\
\label{nl_ch_3_mod_fr_1_newvar}
\tilde E^{(31;1)}_1&:=&
4 v_{t_3} -v_{xxx} - 3 v_{xt_2} -
6 p_{t_2} - 6 \Big(q_{t_2}+ v_{t_2} v  + (v_x)^2+
 \nu_0 v_{t_2}  \Big)v -\\\nonumber
&&
6 v_x\Big(q_x +v_{xy}+v v_x +\nu_0 v_x\Big)-
6 v_{t_2} (q+ v_y) -6 q_x v_x
 =0,
\\\label{nl4_fr_1_newvar}
\tilde E^{(1;1)}_1&:=&v_{yy} + 2  q_y -  v_x +\nu_0 v_y  +
2   v_y  v =0.
\end{eqnarray}
Any two equations out of the system (\ref{nl_ch_1_fr_1_newvar}-\ref{nl_ch_3_mod_fr_1_newvar}) supplemented by eq.(\ref{nl4_fr_1_newvar}) represent the complete system of matrix nonlinear PDEs for the fields $v$, $p$ and $q$. Two examples of the scalar nonlinear PDEs are given below in Sec.\ref{Section:examples}.

%%%%%%%%%%
\subsection{Examples of scalar nonlinear PDEs}
\label{Section:examples}
\paragraph{Example 1.} 
Consider the system (\ref{nl_ch_1_fr_1_newvar},\ref{nl_ch_2_fr_1_newvar},\ref{nl4_fr_1_newvar}). 
In the scalar case this system reduces to the following  single scalar nonlinear PDE for the field $v$:
\begin{eqnarray}
\left((E^{(21;1)}_1)_x-\frac{1}{2}(E^{(11;1)}_1)_y\right)_y + \frac{1}{2}(\tilde E^{(1;1)}_1)_{xx}=0\;\;\;\Rightarrow\;\; {\mbox{eq.}}(\ref{nl_ex1}).
\end{eqnarray}

%%%%%%%%%%%%%%%%%
\paragraph{Example 2.}
Consider the system (\ref{nl_ch_1_fr_1_newvar},\ref{nl_ch_3_mod_fr_1_newvar},\ref{nl4_fr_1_newvar}).
In the scalar case this system reduces  to the single equation for the field $v$   as follows:  
 \begin{eqnarray}
\Big(\frac{1}{4} \tilde E^{(31;1)}_1 +\frac{3 v_{x}}{2}\partial_y^{-1} (\tilde E^{(1;1)}_1)_x\Big)_x-\frac{3}{4}(E^{(11;1)}_1)_{t_2} \;\;\Rightarrow\;\;{\mbox{eq.}}(\ref{nl_ex2}).
 \end{eqnarray}

%In particular, assume that field $r$ does not  depend on $t_2$. %Then he eq.(\ref{nl_ex2}) reduces to the following one:
%\begin{eqnarray}
%v_{t_3}-\frac{1}{4} v_{xxx} -\frac{3}{2} v_x \partial_y^{-1} %v_{xx} =0. 
%\end{eqnarray}

%%%%%%%%%%%%
\section{Solutions to the derived nonlinear PDEs
% derived in Secs.\ref{Section:diff_reduction} and 
%\ref{Section:Frobenius}
 }
\label{Section:solutions}

%%%%%%%%%%%%%%%%%
\subsection{Solutions to  nonlinear system %(\ref{nl_ch_1}-\ref{nl_ch_31},\ref{nl4},\ref{nl4_1})
(\ref{dif_1}-\ref{dif_3})}
Solutions to  systems of matrix  nonlinear PDEs (\ref{dif_1}-\ref{dif_3}) or, more general, to  discrete chains  (\ref{nl_ch_1}-\ref{nl_ch_31},\ref{nl4},\ref{nl4_1})
are obtainable in terms of the functions $\Psi^{(m)}$  $m=1,\dots,2N_0$ taken as  solutions of  linear system (\ref{x},\ref{t2}-\ref{t3}):
\begin{eqnarray} \label{Psi_sol0}
\Psi^{(l)}_{\alpha\beta}(\vec{x}) &=&  \sum_{i=1}^2 \int dq\psi^{(l;i)}_{\alpha\beta}(q) e^{q x+
k^{(i)}y +
(q^2-\tilde \Lambda^{(2;l)}_\beta) t_2+
(q^3-\tilde \Lambda^{(3;l)}_\beta) t_3 +k^{(i)} q\tau} ,
\\\nonumber
&&
\alpha,\beta=1,\dots,n_0M,
\;\;l=1,\dots,2 N_0,
\end{eqnarray}
where $\psi^{(l)}_{\alpha\beta}(q)$  are the arbitrary scalar functions of  single  scalar variable,  $\vec x=(x,y,t_2,t_3)$ is the list of all independent variables of the nonlinear PDEs, $k^{(i)}$, $i=1,2$, are  roots of the  characteristics equation assotiated with eq.(\ref{g_r2}) where $a$, $\nu$ and $\mu$ are given by eqs.(\ref{a_nu_mu}):
\begin{eqnarray}\label{char}
&&
k^2  -   q-\nu_0  k -\mu_0=0, \;\;\Rightarrow\\\nonumber
&&
k^{(1)}=\frac{1}{2}\left(\nu_0 +\sqrt{\nu_0^2+4   q+ 4\mu_0}\right),\;\;
k^{(2)}=\frac{1}{2}\left(\nu_0 -\sqrt{\nu_0^2+4  q  + 4\mu_0}\right)
.
\end{eqnarray}
Now, we can use eq.(\ref{b1}) in order to find $v^{(i)}$ and $w^{(i)}$.  In particular, there is a big class of solutions in the form of rational functions of exponents, such as solitary wave solutions. We do not represent their explicite form since eqs.(\ref{dif_1}-\ref{dif_3}) are intermediate equations in our method of solving eqs.(\ref{nl_ex1},\ref{nl_ex2}). 

Note nevertheless that
the simplest nontrivial solution to eqs.(\ref{dif_1}-\ref{dif_3})  corresponds to 
$N_0=2$. Then  eq.(\ref{b1}) reduces to the following four  matrix equations:
\begin{eqnarray}
\Psi^{(m)}_x =
\sum_{i=1}^4 \Big(v^{(i)} \partial_x^{-i+1}\Psi^{(m)}_y +w^{(i)} \partial_x^{-i+1}\Psi^{(m)}\Big) ,\;\;m=1,2,3,4.
\end{eqnarray}
These equations, in general, are uniquely solvable  for the matrix fields $v^{(i)}$ and $w^{(i)}$, $i=1,2$.  

%%%%%%%%
\paragraph{The reductions to the lower dimensional PDEs.}
The described solution space admits the following reductions
\begin{eqnarray}\label{red_sol_1}
v^{(i)}_{t_n}=w^{(i)}_{t_n}=0,\;\;\forall \;i,
\end{eqnarray}
where $n$ is either 2 or 3.
In fact, these reductions mean 
\begin{eqnarray}
\psi^{(l;i)}_{\alpha\beta}(q)\sim
\delta\left(q^n - \tilde\Lambda^{(n;l)}_\beta\right) 
\end{eqnarray}
in eq.(\ref{Psi_sol0}).
For instance, let $n=2$. Then we must substitute the following expression into eq.(\ref{Psi_sol0}):
 \begin{eqnarray}\label{psi_red} \psi^{(l;i)}_{\alpha\beta}(q)=\psi^{(l;i1)}_{\alpha\beta}
 \delta\left(q-\sqrt{\tilde\Lambda^{(n;l)}_\beta}\right)+
 \psi^{(l;i2)}_{\alpha\beta}
 \delta\left(q+\sqrt{\tilde\Lambda^{(n;l)}_\beta}\right),
 \end{eqnarray}
 where $\psi^{(l;ik)}_{\alpha\beta}$, $k=1,2$, are arbitrary constants.

 %%%%%%%%%%%
 \subsection{Solutions to  system (\ref{nl_ch_1_fr_1_newvar}-\ref{nl4_fr_1_newvar})  and to  scalar equations (\ref{nl_ex1},\ref{nl_ex2}) and (\ref{ex_3dim1},\ref{ex_3dim2})}  

In this case  the fields $v^{(i;j)}$ and $w^{(i;j)}$   are solutions to  linear algebraic system (\ref{Psi_ex}). More precisely, only part of  system  (\ref{Psi_ex}) is needed to define $v^{(i;j)}$ and $w^{(i;j)}$. In fact, system (\ref{Psi_ex}) may be viewed as  two subsystems.
The first one corresponds to $n=1$:
\begin{eqnarray}\label{Psi1}
\Psi^{(l;1m)}_x&=&
\sum_{i=1}^{N_0}\sum_{j=1}^{n_0}\left[v^{(i;j)} 
\partial_x^{-i+1}\Psi^{(l;jm)}_y +
\ w^{(i;j)} 
\partial_x^{-i+1}\Psi^{(l;jm)}\right],\\\nonumber
&&
l=1,\dots,2 N_0\;\;,\;\;m=1,\dots,n_0.
\end{eqnarray}
This is the system of $2 N_0 n_0$  linear algebraic $M\times M$ matrix equations for the same number of the matrix  fields $v^{(i;l)}$ and $w^{(i;l)}$, $i=1,\dots,N_0$, $j=1,\dots,n_0$. Namely eqs.(\ref{Psi1})  yield the functions $v^{(1;1)}$, $w^{(1;1)}$, $v^{(2;1)}$, $v^{(1;1+n_1(1))}$, $w^{(1;1+n_1(1))}$, $v^{(1;1+n_2(1))}$, $v^{(1;1+2n_1(1))}$ defining the solutions $v$, $p$, $q$ to  system (\ref{nl_ch_1_fr_1_newvar}-\ref{nl4_fr_1_newvar}) in accordance with formulae (\ref{newvar}). The
second subsystem corresponds to $n>1$ in (\ref{Psi_ex}):
\begin{eqnarray}\label{Psi_nm}
\Psi^{(l;nm)}_x&=&
\sum_{i=0}^{N_0}\left[
\partial_x^{-i+1}\Psi^{(l;(n-n_1(i))m)}_y +
\partial_x^{-i+1}\Psi^{(l;(n-n_2(i))m)}\right],\\\nonumber
&&l=1,\dots,2 N_0,\;\;n,m=1,\dots,n_0,\;\;\;
\Psi^{(l;ij)}=0,\; {\mbox{if}},\;i\le 0.
\end{eqnarray}
%We may take $n_1(i)=i+1$.
This subsystem  expresses recursively the functions
 $\Psi^{(l;nm)}$, $n>1$, in terms of the functions 
 $\Psi^{(l;1m)}$. The simplest case corresponds to $n_j(i)=1$, $\forall i,j$ (the Frobenius reduction).
 
 The functions $\Psi^{(l;nm)}$  are solutions to  system (\ref{t2_ex}-\ref{x_ex}) and may be written as follows:  
\begin{eqnarray} \label{Psi_sol0_fr}
&&
\Psi^{(l;1m)}_{\alpha\beta}(\vec{x}) =  \sum_{i=1}^2 \int dq\psi^{(lm;i)}_{\alpha\beta}(q) e^{qx+
k^{(i)}y +
(q^2-\Lambda^{(2;l;m)}_\beta) t_2+
(q^3-\Lambda^{(3;l;m)}_\beta) t_3 +k^{(i)} q\tau}
\\\nonumber
&&
\alpha,\beta=1,\dots,M,
\;\;m=1,\dots,n_0,\;\;l=1,\dots,2 N_0,\\\label{k}
&&
k^{(1)}=\frac{1}{2}\left(\nu_0 +\sqrt{\nu_0^2+4  q + 4\mu_0}\right),\;\;
k^{(2)}=\frac{1}{2}\left(\nu_0 -\sqrt{\nu_0^2+4  q + 4\mu_0}\right)
.
\end{eqnarray}
Here $\psi^{(lm;i)}_{\alpha\beta}(q)$  are  arbitrary scalar functions of one scalar  argument. In particular, the derived formulae describe the big class of solutions having the form of rational functions of exponents, such as solitary wave solutions. The simplest examples of them will be given below, see eqs.(\ref{solution}-\ref{kk},\ref{simplest}).

In the scalar case (corresponding to eqs.(\ref{nl_ex1}) and (\ref{nl_ex2}))
one has  $M=1$ so that eq.(\ref{Psi_sol0_fr}) reads:
\begin{eqnarray} \label{Psi_sol0_fr_scalar}
&&
\Psi^{(l;1m)}(\vec{x}) =  \sum_{i=1}^2 \int dq\psi^{(lm;i)}(q) e^{qx+
k^{(i)}y +
(q^2-\Lambda^{(2;l;m)}) t_2+
(q^3-\Lambda^{(3;l;m)}) t_3 +k^{(i)} q\tau}
%\\\label{Lam_sol_ex}\nonumber
%&&
%k^{(1)}=\frac{1}{2}\left(\nu_0 +\sqrt{\nu_0^2+4  q + %4\mu_0}\right),\;\;
%k^{(2)}=\frac{1}{2}\left(\nu_0 -\sqrt{\nu_0^2+4  q + %4\mu_0}\right)
%.
\end{eqnarray}
with the same expressions for $k^{(i)}$ given by eqs.(\ref{k}).

%%%%%%%%
\paragraph{The reductions to the lower dimensional PDEs.}
Note that the described solution space admits reduction (\ref{red}) embedded into the following set of reductions 
\begin{eqnarray}
v^{(i;j)}_{t_n}=w^{(i;j)}_{t_n}=0\;\;\forall \;i,j,
\end{eqnarray}
where $n$ is either 2 or 3.
In fact, similar to reduction (\ref{red_sol_1}), these reductions mean 
\begin{eqnarray}\label{psi_red_fr}
\psi^{(lm;i)}_{\alpha\beta}(q)\sim
\delta\left(q^n - \Lambda^{(n;l;m)}_\beta\right) 
\end{eqnarray}
in eq.(\ref{Psi_sol0_fr}).
% or
%\begin{eqnarray}
%\psi^{(lm;i)}(q)\sim
%\delta\left(q^2 - \Lambda^{(n;l;m)}\right)  
%\end{eqnarray}
% in eq.(\ref{Psi_sol0_scalar}).
For instance, let $n=2$ which corresponds to reduction (\ref{red}). Then we must substitute the following expression into eq.(\ref{Psi_sol0_fr}):
 \begin{eqnarray}\label{psi_red_fr_scalar} \psi^{(lm;i)}_{\alpha\beta}(q)=\psi^{(lm;i1)}_{\alpha\beta}\delta\left(q-\sqrt{\Lambda^{(n;l;m)}_\beta}\right)+
 \psi^{(lm;i2)}_{\alpha\beta}
 \delta\left(q+\sqrt{\Lambda^{(n;l;m)}_\beta}\right),
 \end{eqnarray}
 where $\psi^{(lm;ik)}_{\alpha\beta}$ ($k=1,2$) are arbitrary constants.
 In the scalar case (corresponding to nonlinear equations (\ref{ex_3dim1}) and (\ref{ex_3dim2}))  eq.(\ref{psi_red_fr_scalar}) reads
 \begin{eqnarray}\label{psi_red_scalar} \psi^{(lm;i)}(q)=\psi^{(lm;i1)}\delta\left(q-\sqrt{\Lambda^{(n;l;m)}}\right)+
 \psi^{(lm;i2)}
 \delta\left(q+\sqrt{\Lambda^{(n;l;m)}}\right),
 \end{eqnarray}
 which must be substituted into eq.(\ref{Psi_sol0_fr_scalar}).
 
%%%%%%%%%%%%%%
\paragraph{The simplest  solution to eqs.(\ref{nl_ex1}) and (\ref{nl_ex2}).}
The simplest nontrivial  solution to eqs.(\ref{nl_ex1}) and (\ref{nl_ex2}) corresponds to  $N_0=n_0=1$ and $ \Lambda^{(n;l;m)}=0$, $n=2,3$. Then eq.(\ref{Psi_sol0_fr_scalar}) reads
\begin{eqnarray} \label{Psi_sol0_scalar_simple}
\Psi^{(l;11)}(\vec{x}) &=&  \sum_{i=1}^2 \int dq\psi^{(l1;i)}(q) e^{qx+k^{(l1;i)} y +q^2 t_2+q^3 t_3 +k^{(l1;i)} q \tau },\;\;l=1,2.
\end{eqnarray}
Eqs.(\ref{Psi1}) become the system of  two following equations:
\begin{eqnarray}\label{Psi1_ex}
\Psi^{(l;11)}_x&=&
v^{(1;1)} 
\Psi^{(l;11)}_y +
\ w^{(1;1)} 
\Psi^{(l;11)},\;\;\;
l=1,2 .
\end{eqnarray}
Their solution reads:
\begin{eqnarray}\label{solution}
&&
v\equiv v^{(1;1)}=\frac{\Delta_1}{\Delta},\;\;\Delta=\left|
\begin{array}{cc}
\Psi^{(1;11)}_y &\Psi^{(1;11)}\cr
\Psi^{(2;11)}_y &\Psi^{(2;11)}
\end{array}
\right|,\;\;\Delta_1=\left|
\begin{array}{cc}
\Psi^{(1;11)}_x &\Psi^{(1;11)}\cr
\Psi^{(2;11)}_x &\Psi^{(2;11)}
\end{array}
\right|,\\\label{solution_b}
&&
w^{(1;1)}=\frac{\Delta_2}{\Delta}
,\;\;
\Delta_2=\left|
\begin{array}{cc}
\Psi^{(1;11)}_y &\Psi^{(1;11)}_x \cr
\Psi^{(2;11)}_y &\Psi^{(2;11)}_x 
\end{array}
\right|.
\end{eqnarray}
Formulae (\ref{solution},\ref{solution_b})  have four arbitrary functions of single variable $\psi^{(l1;i)}(q)$, $l,i=1,2$. 

Let us write the explicite formulae  for  the particular case $\psi^{(l1;i)}(q)=\psi_{li}\delta(q-q_{li})$, where both $\psi_{li}$ and $q_{li}$ are arbitrary  constants:
\begin{eqnarray}\label{expl_Delta}
\Delta&=&\sum_{i,j=1}^2 e^{Q_{1i}+Q_{2j}}\Big(
(-1)^{i+1} K_{1i}-(-1)^{j+1} K_{2j}\Big)\psi^{(11;i)}\psi^{(21;j)},\\\label{expl_Delta1}
\Delta_1&=&\sum_{i,j=1}^2 e^{Q_{1i}+Q_{2j}}\Big(
q_{1i}-q_{2j}\Big)\psi^{(11;i)}\psi^{(21;j)},\\\label{expl_Q}
&&
Q_{li}=q_{li}x+k^{(l1;i)} y +q_{li}^2 t_2+q_{li}^3 t_3 +k^{(l1;i)} q \tau,\\\label{kk}
&&
k^{(l1;1)}=\frac{\nu_0}{2} +K_{l1},\;\;
k^{(l1;2)}=\frac{\nu_0}{2} -K_{l2},\;\;
K_{li}=\frac{1}{2}\sqrt{\nu_0^2 + 4 q_{li} + 4\mu_0},\;\;i,l=1,2
.
\end{eqnarray}
The solution $v$ has no singularities if $\Delta\neq 0$, which is true if, for instance, the following relations are valid:
\begin{eqnarray}\label{nonsing}
K_{1i}>K_{2i}>0,\;\;i=1,2,\;\;\psi^{(11;2)}<0,\;\;\psi^{(11;1)},\psi^{(21;1)},\psi^{(21;2)}>0.
\end{eqnarray}
For example, let
\begin{eqnarray}
K_{2i}=q_{2i}=0,\;\;i=1,2\;\;\Rightarrow \;\;\mu_0=-\frac{\nu_0^2}{4},\;\;
\psi_{11}=\xi_1>0,\;\;\psi_{12}=-\xi_2<0
\end{eqnarray}
Then
\begin{eqnarray}\label{simplest}
v&=&\frac{e^{Q_{11}}\xi_1 q_{11} - e^{Q_{12}}\xi_2 q_{12}}{ e^{Q_{11}}q_{11}\xi_1 + e^{Q_{12}}q_{12}\xi_2},
\end{eqnarray}
which is the kink.

%%%%%%%%%%
\paragraph{The simplest solution to  eqs.(\ref{ex_3dim1}) and (\ref{ex_3dim2}).}
We take eq.(\ref{Psi_sol0_fr_scalar}) with $N_0=n_0=1$ and $\Lambda^{(3;l;m)}=0$: 
\begin{eqnarray} \label{Psi_sol0_scalar_simple_red}
\Psi^{(l;11)}(\vec{x}) &=&  \sum_{i=1}^2 \int dq\psi^{(l1;i)}(q) e^{qx+k^{(l1;i)} y +(q^2-\Lambda^{(2;l;m)}) t_2+q^3 t_3 +k^{(l1;i)} q \tau },\;\;l=1,2.
\end{eqnarray}
Note that eqs.(\ref{ex_3dim1}) and
(\ref{ex_3dim2}) require  reduction (\ref{red}), which means
that function $\psi^{(l1;i)}(q)$ in eq.(\ref{Psi_sol0_scalar_simple_red})   is given by 
eq.(\ref{psi_red_scalar}).
For instance, let
\begin{eqnarray}
\psi^{(lm;ij)}=\delta_{ij} \psi_{lmi}.
\end{eqnarray}
Then 
one has expressions
\begin{eqnarray}\label{expl_Delta_red}
\Delta&=&\sum_{i,j=1}^2 e^{Q_{1i}+Q_{2j}}\Big(
(-1)^{i+1} K_{1i}-(-1)^{j+1} K_{2j}\Big)\psi_{11i}\psi_{21j},\\\nonumber
\Delta_1&=&\sum_{i,j=1}^2 e^{Q_{1i}+Q_{2j}}\Big(
(-1)^{i+1}\Lambda^{(11)}-(-1)^{j+1}\Lambda^{(21)}\Big)\psi_{11i}
\psi_{21j},\\\nonumber
&&
Q_{li}=q_{li}x+k^{(l1;i)} y +q_{li}^3 t_3 +k^{(l1;i)} q_{li} \tau,\;\;q_{l1}=\sqrt{\Lambda^{(l1)}},\;\;q_{l2}=-\sqrt{\Lambda^{(l1)}}
%\\\nonumber
%&&
%k^{(l1;1)}=\frac{\nu_0}{2} +K_{l1},\;\;
%k^{(l1;2)}=\frac{\nu_0}{2} -K_{l2},\;\;
%K_{li}=\frac{1}{2}\sqrt{\nu_0^2 + 4 q_{li} + 4\mu_0}
\end{eqnarray}
 instead of expressions (\ref{expl_Delta}-\ref{expl_Q}), while relations (\ref{kk}) remain the same. The
conditions similar to (\ref{nonsing}) with replacement $\psi^{(ij;k)}\to\psi_{ijk}$  guarantee that the  
solution $v$ is non-singular in this case as well.

Let, for example,  $\psi_{212}=0$, $\psi_{111}=\xi_1>0$, $\psi_{112}=-\xi_2<0$ and $K_{11}>K_{21}>0$. Then  one has
\begin{eqnarray}
v&=&\frac{e^{Q_{11}}\xi_1 (q_{11}-q_{21}) + e^{Q_{12}}\xi_2 (q_{11}+q_{21})}{ e^{Q_{11}}(K_{11}-K_{21})\xi_1 + e^{Q_{12}}(K_{11}+K_{21})\xi_2},
\end{eqnarray}
which is the kink.

%%%%%%%%%
\section{Obstacles to the complete integrability of eqs.(\ref{nl_ex1_sc}) and (\ref{nl_ex2})}
\label{Section:obstacle}
It is important to note that our algorithm  does not describe the full solution spaces to  nonlinear PDEs (\ref{nl_ex1_sc}),  (\ref{nl_ex2}), (\ref{ex_3dim1}) and (\ref{ex_3dim2}). To justify this statement we consider eqs.(\ref{nl_ex1_sc}) and (\ref{nl_ex2}), while eqs.(\ref{ex_3dim1}) and (\ref{ex_3dim2}) may be treated in a similar way. The simplest argument is  following. By construction,  eqs.(\ref{nl_ex1_sc}) and (\ref{nl_ex2}) must be commuting flows, i.e. the expression 
\begin{eqnarray}
[{\mbox{eq.(\ref{nl_ex1})}}]_{t_3} -[{\mbox{eq.(\ref{nl_ex2})}}]_{y\tau}
\end{eqnarray}
must be identical to zero. However, the direct calculation  shows that this expression  yields the nonlocal three-dimensional PDE  having  rather complicated form (we do not represent it here). This means that the constructed solution space to the equations (\ref{nl_ex1_sc}) and (\ref{nl_ex2}) is two-dimensional (in other words, it may depend on the arbitrary functions of two independent variables) while the full solution space to four-dimensional PDEs (\ref{nl_ex1_sc}) and (\ref{nl_ex2}) must be three dimensional.
The dimensionality of the solution space is confirmed in Sec.\ref{Section:solutions} by formulae (\ref{Psi_sol0_fr_scalar}), which shows us that the solution space depends on $4 N_0 n_0$ arbitrary functions of single variable $\psi^{(lm;i)}(q)$, $l=1,\dots,2 N_0$, $m=1,\dots,n_0$, $i=1,2$. Since both $N_0$ and $n_0$ are arbitrary positive integers and may go to infinity, these functions  may approximate arbitrary functions of two variables in the solution space. Thus, the possibility to have an arbitrary function of three independent variables in the solution space (which would provide the full integrability) remains an open problem for further study. Without details, we remark that solution space to four-dimensional  eq.(\ref{nl_ex1}) (available due to  our algorithm)  is two-dimensional as well, while solution spaces to three-dimensional equations (\ref{ex_3dim1}) and (\ref{ex_3dim2})
are one-dimensional, which may be obtained owing to the formulae (\ref{Psi_sol0_fr_scalar}) and (\ref{psi_red_scalar}).

Since eqs.(\ref{nl_ex1_sc}) and (\ref{nl_ex2}) have been derived from the matrix KP and the later is well known (2+1)-dimensional PDE integrable by the ISTM, one can expect the same type of integrability of eqs.(\ref{nl_ex1_sc}) and (\ref{nl_ex2}). In particular, one can expect that these equations are compatibility conditions of some overdetermined linear spectral problem derivable  from the  overdetermined linear spectral problem for the matrix KP. However, this is not true. 
In the next subsection we consider eq.(\ref{nl_ex2}) as an example and show that the linear spectral problem for this equation  is not well defined. Thus, at the moment, the ISTM is not a suitable tool for solving eq.(\ref{nl_ex2}). The same conclusion is valid for  eq.(\ref{nl_ex1_sc}), (\ref{ex_3dim1}) and  (\ref{ex_3dim2})  as well.

%%%%%%%%%%%%
\subsection{The spectral problem for the matrix KP supplemented by  differential  reduction (\ref{ww_i},\ref{qp_i}) and Frobenious type reduction (\ref{G_F})}
Although the algorithm described above allows one to find a big manifold of solutions to eqs.(\ref{nl_ex1_sc}), (\ref{nl_ex2}), (\ref{ex_3dim1}) and (\ref{ex_3dim2}) it does not gives us an algorithm to derive the linear spectral problem for any of these equations, i.e., at the moment, we are not able to obtain such overdetermined system of linear equations for some spectral function whose compatibility condition results in eq.(\ref{nl_ex1_sc}), or eq.(\ref{nl_ex2}), or eq.(\ref{ex_3dim1}), or eq.(\ref{ex_3dim2}) without additional requirements to the coefficients of linear system.
%(unlike the algorithm based on the single Frobenious reduction %\cite{ZS_2}).
 As an example explaining this problem,  we consider eq.(\ref{nl_ex2}) as nonlinear equation derivable from  the matrix KP (\ref{matrixKP}) after differential reduction (introduced in Sec.\ref{Section:diff_reduction}) followed by the Frobenious type reduction (introduced in Sec.\ref{Section:FTR}) in this subsection.
 
The obstacle to derive the linear spectral problem for eq.(\ref{nl_ex2}) is assotiated with the spectral representation of differential reduction (\ref{ww_i},\ref{qp_i}) which is a reduction of the potentials of the linear spectral problem for the matrix KP. In other words, we do not know which reduction must be imposed on the spectral function  to generate the above differential reduction for the potentials of the spectral problem.  Thus the problem of spectral representation 
of the eq.(\ref{nl_ex2}) remains open as well as the problem of its complete integrability. Nevertheless, we derive some linear spectral problem whose compatibility condition leads to  eq.(\ref{nl_ex2}) imposing the  above differential reduction "by hand".

The spectral problem for  matrix KP (\ref{matrixKP}) reads
\begin{eqnarray}\label{sp_p_KP}
F^{(1)}&:=&\psi_{t_2}(\lambda;\vec x)+ \psi_{xx}(\lambda;\vec x) +2 \psi(\lambda;\vec x) W^{(1)}_x(\vec x)=0,\\\nonumber
F^{(2)}&:=&\psi_{t_3}(\lambda;\vec x)- \psi_{xxx}(\lambda;\vec x) -
3  \psi_{x}(\lambda;\vec x) W^{(1)}_x(\vec x)+\frac{3}{2} \psi(\lambda;\vec x)\left(
W^{(1)}_{t_2}(\vec x)-W^{(1)}_{xx}(\vec x)
\right)=0,
\end{eqnarray}
where $\psi$ is a spectral function and $\lambda$ is a spectral parameter.
Differential reduction (\ref{ww_i},\ref{qp_i})  requires the following block structure of $\psi$:
\begin{eqnarray}
\psi=\left(
\begin{array}{cc}
\psi^{(11)} &\psi^{(12)} \cr
\psi^{(21)} &\psi^{(22)}
\end{array}\right),
\end{eqnarray}
where $\psi^{(ij)}$ are $n_0 M \times n_0M$ matrices. 
So that equations (\ref{sp_p_KP}) acquire the following block structure:
\begin{eqnarray}\label{F_red}
F^{(i)}\equiv\left(\begin{array}{cc}
F^{(i;11)}&F^{(i;12)}\cr
F^{(i;21)}&F^{(i;22)}
\end{array}\right)=0,\;\;i=1,2
\end{eqnarray}
where $F^{(i;nm)}$ are $n_0M\times n_0 M$ matrix equations.
The  spectral problem is now represented by the first-row blocks  of eq.(\ref{F_red}):
\begin{eqnarray}\label{lin_dif_1}
F^{(1;11)}&:=& \psi^{(11)}_{t_2} + \psi^{(11)}_{xx} + 2  \psi^{(11)} w^{(1)}_x +
 2  \psi^{(12)} p^{(1)}_x=0 
,\\\label{lin_dif_2}
F^{(1;12)}&:=&\psi^{(12)}_{t_2} + \psi^{(12)}_{xx} + 2  \psi^{(11)} v^{(1)}_x +
 2  \psi^{(12)} q^{(1)}_x=0 ,\\\label{lin_dif_3}
F^{(2;11)}&:=& \psi^{(11)}_{t_3} - \psi^{(11)}_{xxx} -3  \left(\psi^{(11)}_x w^{(1)}_x +
   \psi^{(12)}_x p^{(1)}_x\right)+\\\nonumber
&&
\frac{3}{2} \left( \psi^{(11)}  
\left(w^{(1)}_{t_2}-w^{(1)}_{xx}\right) +\psi^{(12)}  
\left(p^{(1)}_{t_2}-p^{(1)}_{xx}\right)\right)=0 
,\\\label{lin_dif_4}
F^{(2;12)}&:=& \psi^{(12)}_{t_3} - \psi^{(12)}_{xxx} -3 \left( \psi^{(11)}_x v^{(1)}_x +
   \psi^{(12)}_x q^{(1)}_x\right)+\\\nonumber
&&\frac{3}{2} \left( \psi^{(11)}  
\left(v^{(1)}_{t_2}-v^{(1)}_{xx}\right) +\psi^{(12)}  
\left(q^{(1)}_{t_2}-q^{(1)}_{xx}\right)\right)=0 ,
\end{eqnarray}
while the second-row blocks of eq.(\ref{F_red})  are equivalent to the system (\ref{lin_dif_1}-\ref{lin_dif_4}) up to the replacement of the spectral functions $\psi^{(1k)}\to\psi^{(2k)}$, $k=1,2$. Emphasize that the differential reduction is already imposed on the coefficients of the above linear system due to the special form of the functions $p^{(1)}$ and $q^{(1)}$ given by eq.(\ref{qp_i}).
It may be shown by the direct calculations that the compatibility conditions of  eqs.(\ref{lin_dif_1}-\ref{lin_dif_4}) 
%supplemented by  equations (\ref{nl4},\ref{nl4_1}) 
yield the system of nonlinear PDEs which is equivalent to  system (\ref{dif_1}), (\ref{dif_3}), i.e. we derive two complete compatible systems of nonlinear PDEs so that eqs.(\ref{dif_1}) describe the $t_2$-evolution  while  eqs.(\ref{dif_3}) describe the $t_3$-evolution of fields. 
However, since differential reduction has not spectral representation, these nonlinear systems may not be integrated by the ISTM.  Of course, this means that the eq.(\ref{nl_ex2})  (which has been derived from the above system of nonlinear PDEs after the Frobenious type reduction (\ref{G_F})) may not be integrated by the ISTM as well. However, for the sake of completeness, we consider the  Frobenious type reduction (\ref{G_F}) imposed on the linear  system (\ref{lin_dif_1}-\ref{lin_dif_4}). 

First of all, note that this reduction requires the following structure of the functions $\psi^{(11)}$ and $\psi^{(12)}$: 
\begin{eqnarray}
\psi^{(1n)}=\left(
\begin{array}{ccc}
\psi^{(n;11)}&\cdots & \psi^{(n;1n_0)}\cr
\cdots&\cdots&\cdots\cr
\psi^{(n;n_01)}&\cdots & \psi^{(n;n_0n_0)}
\end{array}
\right),\;\;n=1,2,
\end{eqnarray}
where  $\psi^{(n;ij)}$ are $M\times M$  functions.
Eqs.(\ref{lin_dif_1}-\ref{lin_dif_4}) acquire the following block structure:
\begin{eqnarray}
F^{(n;1m)}=\left(
\begin{array}{ccc}
F^{(nm;11)} &\cdots&F^{(nm;1n_0)}\cr
\cdots&\cdots&\cdots\cr
F^{(nm;n_01)} &\cdots&F^{(nm;n_0n_0)}
\end{array}
\right),\;\;n,m=1,2,
\end{eqnarray}
where $F^{(nm;ij)}$ are $M\times M$ matrix  equations. Similarly to eq.(\ref{F_red}), only the first-row blocks of these equations represent the system of independent spectral equations:
\begin{eqnarray}\label{lin_dif_Fr_1}
F^{(11;n)}&:=& \psi^{(1;n)}_{t_2} + \psi^{(1;n)}_{xx} + 2  \psi^{(1;1)} w^{(1;n)}_x +
 2  \psi^{(2;1)} p^{(1;1n)}_x+2  \psi^{(2;2)} p^{(1;2n)}_x=0 
,\\\label{lin_dif_Fr_2}
F^{(12;n)}&:=&\psi^{(2;n)}_{t_2} + \psi^{(2;n)}_{xx} + 2  \psi^{(1;1)} v^{(1;n)}_x +
 2  \psi^{(2;1)} q^{(1;1n)}_x+2  \psi^{(2;2)} q^{(1;2n)}_x=0 ,\\\label{lin_dif_Fr_3}
F^{(21;n)}&:=& \psi^{(1;n)}_{t_3} - \psi^{(1;n)}_{xxx} -3  \left(\psi^{(1;1)}_x w^{(1;n)}_x +
   \psi^{(2;1)}_x p^{(1;1n)}_x+ \psi^{(2;2)}_x p^{(1;2n)}_x\right)+\\\nonumber
&&\hspace{-1cm}
\frac{3}{2} \left( \psi^{(1;1)}  
\left(w^{(1;n)}_{t_2}-w^{(1;n)}_{xx}\right) +\psi^{(2;1)}  
\left(p^{(1;1n)}_{t_2}-p^{(1;1n)}_{xx}\right)+\psi^{(2;2)}  
\left(p^{(1;2n)}_{t_2}-p^{(1;2n)}_{xx}\right)\right)=0 
,\\\label{lin_dif_Fr_4}
F^{(22;n)}&:=& \psi^{(2;n)}_{t_3} - \psi^{(2;n)}_{xxx} -3 \left( \psi^{(1;1)}_x v^{(1;n)}_x +
   \psi^{(2;1)}_x q^{(1;1n)}_x+\psi^{(2;2)}_x q^{(1;2n)}_x\right)+\\\nonumber
&&\hspace{-1cm}
\frac{3}{2} \left( \psi^{(1;1)}  
\left(v^{(1;n)}_{t_2}-v^{(1;n)}_{xx}\right) +\psi^{(2;1)}  
\left(q^{(1;1n)}_{t_2}-q^{(1;1n)}_{xx}\right)+\psi^{(2;2)}  
\left(q^{(1;2n)}_{t_2}-q^{(1;2n)}_{xx}\right)\right)=0,
\end{eqnarray}
where $n=1,\dots,n_0$ and
\begin{eqnarray}
&&
p^{(1;1k)}=w^{(1;k)}_y + \mu_0 v^{(1;k)} + v^{(2;k)} + v^{(1;1)} w^{(1;k)}+
   v^{(1;k+n_2(1))},\;\;p^{(1;2k)}=w^{(1;k)},\\\nonumber
&&
q^{(1;1k)}=v^{(1;k)}_y + \nu_0 v^{(1;k)} + w^{(1;k)} + v^{(1;1)} v^{(1;k)}+
   v^{(1;k+1)},\;\;p^{(1;2k)}=v^{(1;k)}.
\end{eqnarray}
%This system must be supplemented by 
%eqs.( \ref{nl4_fr},\ref{nl4_1_fr}), $i=l=1$ which appear after 
%the Frobenious type reduction imposed on 
%eqs.(\ref{nl4},\ref{nl4_1}). 
Here we take into account eq.(\ref{a_nu_mu})
 and take $n_1(1)=1$ without loss of generality.
The  system of linear PDEs (\ref{lin_dif_Fr_1}-\ref{lin_dif_Fr_4}), $n=1,2$ represents the overdetermined system  whose compatibility condition  results in  the complete system of nonlinear PDEs which includes
 equations (\ref{nl_ch_1_fr_1_newvar}), (\ref{nl_ch_3_mod_fr_1_newvar}) and 
 (\ref{nl4_fr_1_newvar}) 
 as a complete subsystem, whose
 scalar version ($M=1$) results in 
eq.(\ref{nl_ex2}).

%%%%%%%%
\section{Conclusions}
\label{Section:conclusions}
We have constructed two examples of the four-dimensional  nonlinear PDEs starting with the dressing method for the (1+1)-dimensional $C$-integrable B\"urgers hierarchy and using combination of the Frobenius type and  the differential reductions. One of these examples, eq.(\ref{nl_ex2}), has the same dispersion relation as the KP does  and may be referred to as the KP type equation. As a consequence, its lower dimensional reduction, eq.(\ref{ex_3dim2}), has the same dispersion relation as the KdV does and may be referred to as the KdV type equation.
Although the derived four-dimensional PDEs are not completely integrable by our technique (see Sec.\ref{Section:obstacle}), we are able to supply a big solution manifold to them with solitary wave solutions as most simple examples.  The new feature of this algorithm in comparison with one represented in \cite{ZS_2} is the differential reduction which introduces the new independent variable $y$ into the nonlinear PDEs. In turn, this variable allows one to introduce the  set of new $\tau$-variables by formula (\ref{tau}), which is a new method of increasing the dimensionality of solvable nonlinear PDEs. 
An important problem is to overcome the obstacle to the complete integrability of the derived four-dimensional nonlinear PDEs (\ref{nl_ex1},\ref{nl_ex2}) and their three dimensional reductions (\ref{ex_3dim1},\ref{ex_3dim2}). 
It is also important to find the physical application of the derived nonlinear PDEs. Regarding this problem we must note that   the additional independent variable $y$ appears only in the nonlinear terms of eqs.(\ref{nl_ex2}) and (\ref{ex_3dim2}),  which usually means the existence of solutions with  break of wave profiles. So that one can expect that these equations describe the break of wave profiles in the physical systems where the KP and the KdV appear. 

The work  was supported by the RFBR grants 
07-01-00446, 08-01-90104 and 09-01-92439
%Mol 08-01-90104
%italy 09-01-92439,
 and by the grant NS-4887.2008.2.

%%%%

%%%%%%%%%%%%%%%%%%%%%%%%%%%%%%%%%%
%%%%%%%%%%%%%%%%%%%%%%%%%%%%%%%%%


\begin{thebibliography}{99}
%%%%%%%%%%%%%%%%%%%%%%%%%%%%%%%%%%
%%%%%%%%%%%%%%%%%%%%%%%%%%%%%%%%%%
\bibitem{GGKM}
C.S.Gardner, J.M.Green, M.D.Kruskal, R.M.Miura, Phys.Rev.Lett,
{\bf 19},  (1967) 1095



\bibitem{ZSh1}
V.E.Zakharov and A.B.Shabat, Funct.Anal.Appl., {\bf 8}, (1974) 43 

\bibitem{ZSh2}
V.E.Zakharov and A.B.Shabat, Funct.Anal.Appl., {\bf 13},  (1979) 13 


\bibitem{OSTT}
Yu.Ohta, Ju.Satsuma, D.Takahashi and T.Tokihiro,
Progr.Theor.Phys.Suppl., No.94 (1988) 210


\bibitem{ts1}
S.P.Tsarev, 
{\it Soviet Math. Dokl.}, {\bf 31},  n. 3, (1985) 488

\bibitem{dn}
B.A. Dubrovin and S.P. Novikov,
{\it Russian Math. Survey}, {\bf 44} n.6 (1989) 35


\bibitem{Ward}
R. S. Ward, Phys. Lett. {\bf 61A} (1977) 81-82

\bibitem{BZ}
A. A. Belavin and V. E. Zakharov, Phys. Lett. {\bf 73B} (1978) 53-57


\bibitem{Krichever}
I. M. Krichever, Comm. Pure Appl. Math. {\bf 47}, 437-475 (1994).

\bibitem{TT}
 K. Takasaki and T. Takebe, Rev. Math. Phys. {\bf 7}, 743 (1995).


\bibitem{DMT}
M. Dunajski, L. J. Mason and P. Todd, J. Geom. Phys. {\bf 37} 63-93 (2001).

\bibitem{KAR}
 B. Konopelchenko, L. Martinez Alonso and O. Ragnisco,  J.Phys. A: Math. Gen. {\bf 34} 10209-10217 (2001).

%\bibitem{GMA}
% F. Guil, M. Manas and L. Martinez Alonso, \u201cOn twistor %solutions of the
% dKP equation\u201d, arXiv:nlin.SI/0211020.


\bibitem{MS1}
S.V.Manakov and P.M.Santini, Phys. Lett. A {\bf 359}, (2006) 613

\bibitem{MS2}
S. V. Manakov and P. M. Santini, JETP Letters, {\bf 83}, No 10, 462-466 (2006)


\bibitem{Z5}
A.I.Zenchuk, J. Math. Phys. V. 49, 063502 (2008)
%Lower-dimensional reductions of GL(M,C) self-dual Yang Mills %equation: Solutions with break of wave profiles
% arXiv:0708.2050v1 [nlin.SI] 


\bibitem{ZS_2}
 A.I.Zenchuk and P.M.Santini,
 J.Phys.A:Math.Theor., V.41 (2008) 185209
%{\it On the remarkable relations among PDEs integrable by 
%the inverse spectral transform method, by
%the method of characteristics and by the Hopf-Cole %transformation}, 
 %arXiv:0801.3945v1 [nlin.SI]



\bibitem{Hopf} E. Hopf, Commun. Pure Appl. Math. {\bf 3}, 201 (1950). 
J. D. Cole, Quan. Appl. Math. {\bf 9}, 225 (1951).



\bibitem{Calogero}
F.Calogero in {\it What is integrability} by V.E.Zakharov,
Springer,  (1990) 1




\bibitem{Calogero2}
 Calogero F and Xiaoda Ji 1991 J. Math. Phys. 32 875

\bibitem{Calogero3}
 Calogero F and Xiaoda Ji 1991 J. Math. Phys. 32 2703

\bibitem{Calogero4}
Calogero F 1992 J. Math. Phys. 33 1257

\bibitem{Calogero5}
 Calogero F 1993 J. Math. Phys. 34 3197

\bibitem{Calogero6}
 Calogero F and Xiaoda Ji 1993 J. Math. Phys. 34 5810

\bibitem{Whitham} J. B. Whitham, {\it Linear and Nonlinear Waves}, Wiley, NY, 1974

\bibitem{SZ} P. M. Santini and A. I. Zenchuk,
%: ``The general solution of the matrix equation 
%$w_t+\sum\limits_{k=1}^nw_{x_k}\rho^{(k)}(w)=\rho(w)+[w,T\tilde\%rho(w)]$''; 
Phys.Lett.A
{\bf  368} (2007) 48
%-52, arXiv:nlin.SI/0612036



\bibitem{ZMNP}
V.E.Zakharov, S.V.Manakov, S.P.Novikov and L.P.Pitaevsky, 
{\it Theory of Solitons. The Inverse Problem Method}, Plenum Press (1984)

\bibitem{CD} F. Calogero and A. Degasperis, {\it Spectral transform and solitons : tools to solve and 
investigate nonlinear evolution equations}, North-Holland, Amsterdam (1982).

\bibitem{AC}
M.J.Ablowitz and P.C.Clarkson, {\it Solitons, Nonlinear Evolution Equations and Inverse Scattering}, 
Cambridge University Press, Cambridge, 1991

\bibitem{Konop}
B. Konopelchenko, {\it Solitons in Multidimensions}, World Scientific, Singapore (1993)


\bibitem{KdV} 
D. J. Korteweg and G. de Vries, Philos. Mag. Ser. 5, {\bf 39},
(1895) 422 




\bibitem{ZS_NLS}
V. E. Zakharov and A. B. Shabat, Sov. Phys. JETP {\bf 34} (1972) 62-69

\bibitem{KP}
B. B. Kadomtsev and V. I. Petviashvili, Sov. Phys. Dokl., {\bf 15} (1970) 539-541

\bibitem{DS}
A. Davey and K. Stewartson, Proc. Roy. Soc. London A, {\bf 338} (1974) 101-110


\bibitem{Z3}
A.I.Zenchuk, to appear in J.Math.Phys., 	arXiv:0901.0647v1 [nlin.SI]
%On the relationship between nonlinear equations integrable by 
%the method of characteristics and equations associated with 
%commuting vector fields

\end{thebibliography}
\end{document}